\documentclass[hidelinks, conference]{IEEEtran}
\makeatletter
\def\ps@headings{%
\def\@oddhead{\mbox{}\scriptsize\rightmark \hfil \thepage}%
\def\@evenhead{\scriptsize\thepage \hfil \leftmark\mbox{}}%
\def\@oddfoot{}%
\def\@evenfoot{}}
\makeatother
\newcommand*{\longDefiningEquals}{\stackrel{\text{def}}{=\joinrel=}}

\usepackage{color}
\usepackage{url}
\usepackage[latin9]{inputenc}
\usepackage{amsthm}
\usepackage{amsmath}
\usepackage{amsthm}
\usepackage{amssymb}
\usepackage{graphicx}
\usepackage{epstopdf}
\usepackage{wrapfig}
\usepackage{algorithmic}
\usepackage{algorithm}
\usepackage{mathrsfs}
\usepackage{float}
\usepackage{mathtools}
\usepackage{tabulary}
\usepackage{booktabs}
\usepackage{caption}
\usepackage{subfig}
\usepackage{xprintlen}
\usepackage{multirow}
\usepackage{textcomp}

\graphicspath{{./}{./files/figures_eps/}}

\PassOptionsToPackage{bookmarks={false}}{hyperref}

\IEEEoverridecommandlockouts
\usepackage{graphicx}
\usepackage{svg}

\def\BibTeX{{\rm B\kern-.05em{\sc i\kern-.025em b}\kern-.08em
   T\kern-.1667em\lower.7ex\hbox{E}\kern-.125emX}}

\newcommand{\comment}[1]{ }

\newcommand\subparagraph{%
  \@startsection{subparagraph}{0}
  {\parindent}
  {0ex \@plus 0ex \@minus 0ex}
  {-1em}
  {\normalfont\normalsize\bfseries}}
\makeatother

\usepackage{titlesec}
\titlespacing*{\section}
{0pt}{1.2ex}{0ex} 
\titlespacing*{\subsection}
{0pt}{1ex}{0ex}  
\titlespacing*{\subsubsection}
{0pt}{0ex}{0ex} 

\begin{document}



\title{On the Domain Generalizability of RF Fingerprints Through Multifractal Dimension Representation}

\author{Benjamin Johnson and Bechir Hamdaoui~\\
School of Electrical Engineering and Computer Science~\\
 Oregon State University, Corvallis, OR, USA ~\\ 
 \{johnsbe3, hamdaoui\}@oregonstate.edu
 \thanks{This work is supported in part by NSF/Intel Award No. 2003273.}
}

\maketitle

\begin{abstract}
RF data-driven device fingerprinting through the use of deep learning has recently surfaced as a possible method for enabling secure device identification and authentication. Traditional approaches are commonly susceptible to the domain adaptation problem where a model trained on data collected under one domain performs badly when tested on data collected under a different domain. Some examples of a domain change include varying the location or environment of the device and varying the time or day of the data collection. In this work, we propose using multifractal analysis and the variance fractal dimension trajectory (VFDT) as a data representation input to the deep neural network to extract device fingerprints that are domain generalizable. We analyze the effectiveness of the proposed VFDT representation in detecting device-specific signatures from hardware-impaired IQ (in-phase and quadrature) signals, and we evaluate its robustness in real-world settings, using an experimental testbed of 30 WiFi-enabled Pycom devices. Our experimental results show that the proposed VFDT representation improves the scalability, robustness and generalizability of the deep learning models significantly compared to when using IQ data samples.
\end{abstract}
 

\section{Introduction}
\label{intro}
Wireless device identification and authentication through the use of RF (radio frequency) fingerprinting has recently been considered as a new potential network security method~\cite{8715341,gul2023secure,shen2023deep}. 
In essence, RF-based device fingerprinting consists of extracting hardware-impaired, device-specific signatures and features from raw RF signals that are transmitted by the devices. These hardware impairments typically come from the inherent variability in the manufacturing process of the devices. The commonly used methods of extracting the fingerprints and classifying the devices mostly rely on deep learning models, which need to be trained first on labeled RF data, and then used to identify and classify devices~\cite{jagannath2022comprehensive,gaskin2022tweak,one_DeepRadioID2019}.

\begin{figure}[t]
\centering
\includegraphics[width=0.95\columnwidth]{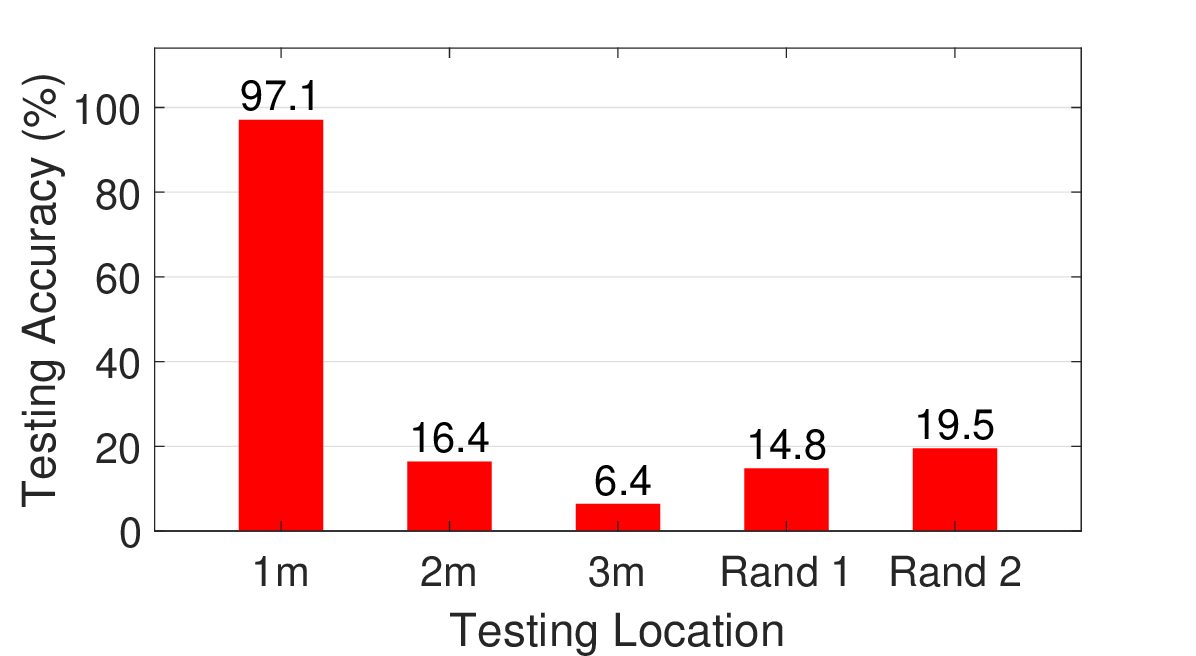}
\caption{CNN model classifier trained on data collected from 30 devices placed 1m away from the receiver, but tested on data collected when the devices are placed at different distances from the receiver: 1m, 2m, 3m, and two random locations. Experiments are taken indoor, in a lab environment.}
\label{fig:motivation}
\end{figure}

While deep learning models offer promising results, they come with some inherent limitations and problems~\cite{elmaghbub2021Lora}. 
Due to the design nature of these learning models and how deep neural networks are trained, the exact features being used to identify and distinguish devices are unknown. This essentially means the deep learning model is a black box and as such could be using something other than the inherent hardware variations to classify devices. For instance, these models could be focusing on the channel conditions instead of the hardware impairments which would lead to incorrect classification when the model is used under different channel conditions. This limitation of the models not being able to generalize to other conditions and settings is often referred to, in the fingerprinting community, as the {\em domain adaptation} or {\em model generalization} problems. 

It has been shown in several studies (e.g.~\cite{gaskin2022tweak,elmaghbub2021Lora,two_deepLoRa2021,three_receiverAgnostic2019,four_scalableLoRa2022}) that deep learning models are able to classify devices with very high accuracy when both the testing data and training data are taken under the same domain (e.g., same channel, time, receiver, location). However, when these models are tested on data taken under a different domain (e.g., training and testing data are collected under different channel conditions), their accuracy is greatly reduced. To demonstrate this domain-adaption challenge, we collected an experimental RF dataset using our testbed, consisting of 30 Pycom devices (more details are provided in Sec.~\ref{sec:testbed}) and fed it to a classical CNN (Convolution Neural Networks) classifier while considering varying distances between the transmitters and the receiver, thereby varying the wireless channel.
The results of this experiment displayed in Fig.~\ref{fig:motivation} show the testing accuracy of the deep learning model, trained on data collected when the devices are placed 1m away from the receiver but tested on data collected when the devices are placed at different locations, 1m, 2m, 3m and at random.
These results show a significant drop in accuracy when the model is tested on data taken from a different domain, in this case the physical location. Note that when the training and testing are both done when the devices are 1m away from the receiver, the testing accuracy is above 97\%. However, due to the limitation mentioned above; i.e., the inability of the learning model to adapt to domain changes, when the model is trained on data collected 1m away but tested on data collected 2m away, the testing accuracy drops from 97\% to only about 16\%.

There have been some attempts that aimed to address the domain adaptation problem in RF fingerprinting \cite{joo2020hold,9732439,gaskin2022tweak,four_scalableLoRa2022,elmaghbub2021Lora,3568100,elmaghbub2020leveraging}. Some works looked at ways to preprocess data to extract specific hardware features, such as the carrier frequency offsets~\cite{joo2020hold}. Other attempts used various custom types of preprocessing to help extract a device-specific, domain-independent fingerprints, which are then passed through a deep learning model for classification \cite{9732439,four_scalableLoRa2022}. There have also been attempts to directly calibrate the learning model itself through few-shot samples to improve its portability across domains \cite{gaskin2022tweak}. Another potential approach~\cite{10012578} relies on multifractal analysis to extract device impairments and use them as features for wireless device identification. 
However, this approach does not examine how the obtained identification results are affected by domain adaptation, nor does it assess how scalable it is with the number of devices to be identified. 

Aside from the domain-adaptation related problems, other works have tackled other device fingerprinting issues, including scalability \cite{9771750} and signal interference \cite{3529520}. For instance, the authors in \cite{10012578} propose using a more complex device fingerprint by combining several different features, including the use of fractal dimension analysis, to use in deep learning models. While they considered changes in basic channel noise, adaptation to other domains has not been explored. Unlike these prior approaches, our work proposes to leverage multifractal signal analysis to extract device fingerprints that are both scalable and robust to domain changes.

This paper proposes using multifractal analysis to extract the inherent hardware-impaired device features from received RF signals and present it as an input to the deep learning classifiers. Specifically, the proposed method involves capturing and sampling the received RF signals, and separately calculating the variance fractal dimension trajectory (VFDT) of both the in-phase (I) and quadrature (Q) components. The resulting VFDT output signals are then fed into a CNN-based deep learning model for device classification. 
We begin by analyzing the effectiveness of the proposed VFDT data representation in detecting and capturing device-specific signatures, caused by hardware-impaired distortions in the RF signals, through real RF datasets captured from both LoRa-enabled and WiFi-enabled devices, as well as through simulated data. We then assess the ability of the proposed representation in identifying and classifying wireless devices, and its ability to adapt to changes in the location domain, and we do so using an experimental testbed of 30 WiFi-enabled Pycom devices. 
We show that the proposed VFDT representation enhances the scalability, robustness and generalizability of the deep learning models significantly compared to using IQ data representation.


The remainder of this work is as follows: Sec.~\ref{sec:background} provides some background on multifractal signal analysis. Sec.~\ref{sec:impairments} examines the effect of individual impairments on the behavior of the proposed VFDT representation. Sec.~\ref{sec:proposed} presents our proposed device classification method. Sec.~\ref{sec:testbed} describes the testbed and data collection used to evaluate the proposed method, and Sec.~\ref{sec:results} evaluates and analyzes the effectiveness of our method. Finally, we conclude the paper in Sec.~\ref{sec:conc}.






%


\section{Variance Fractal Dimension Trajectory}
\label{sec:background}
Multifractal analysis uses the fractal dimension to characterize how a signal varies or meanders over different scales of measurements. It has been used in various real-world applications ranging from noise estimation~\cite{117957} to fish trajectory analysis~\cite{1225951} to estimating the length of coastlines \cite{1532616}. In general, the fractal dimension can be seen as a representation of the degree of irregularity, complexity, or meandering  of an object or signal~\cite{multifractal}. There are many different classes of fractal dimensions such as morphological fractal dimensions, entropy-based fractal dimensions, and transform-based fractal dimensions~\cite{1532616}. In this paper, we focus on a specific type of transform-based fractal dimensions, known as the variance fractal dimension. This section shows how this type of signal analysis can be used to extract the hardware-impaired fingerprints of wireless devices from received RF signals.


\subsection{The Variance Fractal Dimension}
Our proposed data representation for enabling efficient fingerprint extraction involves estimating the variance fractal dimension of the RF signals using multifractal analysis. This variance dimension is based on analyzing the statistical variance of the signal amplitude over different scales and ranges. In order to estimate this variance dimension, it is first assumed that the signal, over a time interval $\Delta  t = t_2 - t_1$, adheres to the power law relationship, $\operatorname{var}[\Delta x] \sim |\Delta  t|^{2H}$,
where $\Delta x=x(t_2)-x(t_1)$ with \(x(t)\) being the signal at time $t$ and \(H\) is the Hurst exponent \cite{1532616}. From this power law relationship, it then follows that~\cite{1532616}
\begin{equation}\nonumber
    H = \lim_{\Delta t \to 0} \frac{1}{2} \frac{\log[\operatorname{var}(\Delta x)_{\triangle t}]}{\log(\Delta t)}
\end{equation}
The variance fractal dimension, \(D\), is then related to the Hurst exponent by $D = E + 1 - H$
%
where \(E\) is the Euclidean dimension \cite{1225951}. For the case of RF signals, after the in-phase and quadrature components are separated, we have $E=1$ (one-dimensional) or $D = 2 - H$, where $D$ again represents the fractal dimension of the overall signal.
%
%
%

\subsection{The VFDT Data Representation}
Our observation here is that for the time-varying RF signals, it is a more useful representation to capture the time-variability of the fractal dimension, and as such, we propose to use the Variance Fractal Dimension Trajectory (VFDT), a rolling trajectory of the fractal dimension, to represent the IQ data that is to be fed to the deep learning classifiers. Later in the paper, we show the effectiveness of this representation on addressing the domain-adaption problem of device fingerprinting. But here in this section, we focus our attention to analyzing and understanding how effective VFDT of the IQ signals, as a data representation, is in distinguishing and separating between devices. Before doing so, let's first describe how we go about calculating VFDT of a time-varying signal.
 
To calculate VFDT of a discrete-time signal, the variance fractal dimension, \(D\), is calculated over a windowed segment of the sampled signal, with the segment being first shifted or offset by a fixed, predetermined amount of samples, and then used for calculating the fractal dimension. The process is repeated until the end of the signal samples is reached. Due to the signal being discrete, the length of the windowed segment, $\Delta w$, is used as the time interval \(\Delta t\), and VFDT of a given windowed segment \(i\) of size $\Delta w$ can be estimated as
\begin{equation}
    \operatorname{VFDT}(i) \longDefiningEquals D(i)= 2 - {\log[\operatorname{var}(\Delta x)]}/{(2\log(\Delta w))}
    \label{eg:vfdt}
\end{equation}
The calculated VFDT depends on both the windowed segment length and the window offset value, whose optimal values vary depending on the signal type. For the RF signals being analyzed in this paper, the main factors that resulted in needing to vary the window length and offset value seemed to be the sample rate as well as the stationarity of the signal. Generally, the best window length and window offset value are subjective and commonly determined through experimentation. We found it useful to have the window offset value be smaller than the window length, as this causes an overlap in adjacent windows and typically yields better results. Also note that the smaller the window length and window offset valule are, the larger the resulting VFDT will be in terms of analyzed segments.



\subsection{VFDT Separability Across Different Devices}\label{subsec:sep}
Our objective in this section is to assess VFDT's ability in separating/distinguishing between wireless devices. For this, we used an USRP receiver to collect real IQ signals sent by multiple different Pycom devices, and applied and visualized the VFDT on each of the received IQ signal to see how separable the calculated VFDTs are across the different devices/signals. 
All Pycom devices are enabled with both LoRa and WiFi communication protocols, and as such, we assess both protocols. For more details on the testbed, refer to Sec.~\ref{sec:testbed}. 
 

\subsubsection{LoRa Protocol}
We collected IQ signal data from 10 different Pycom (LoPy) devices, each enabled with the LoRa transmission mode and sent at 915MHz. Data was sampled by the USRP receiver at 1MS/s. Each Pycom device was placed 5m away from the receiver in an indoor environment and the transmitted data was collected for 20s. 
Plots of the calculated VFDT for the I signal component of 3 different Pycom devices are shown in Fig.~\ref{fig:lora}. From this figure, it can be seen that all 3 devices have relatively consistent VFDT values. More importantly, all 3 devices exhibit well separable VFDT values, and thus can be well distinguished from one another. The results also hold for the other tested devices but are not shown in the figure. All tested devices have a unique average value that is consistent throughout the transmission. These experimental results suggest that the combination of different hardware impairments for each device creates a unique fingerprint that is well captured through the proposed VFDT representation.

\subsubsection{WiFi Protocol}
\begin{figure}[t]
\centering
\subfloat[LoRa]{
    \includegraphics[width=0.5\columnwidth]{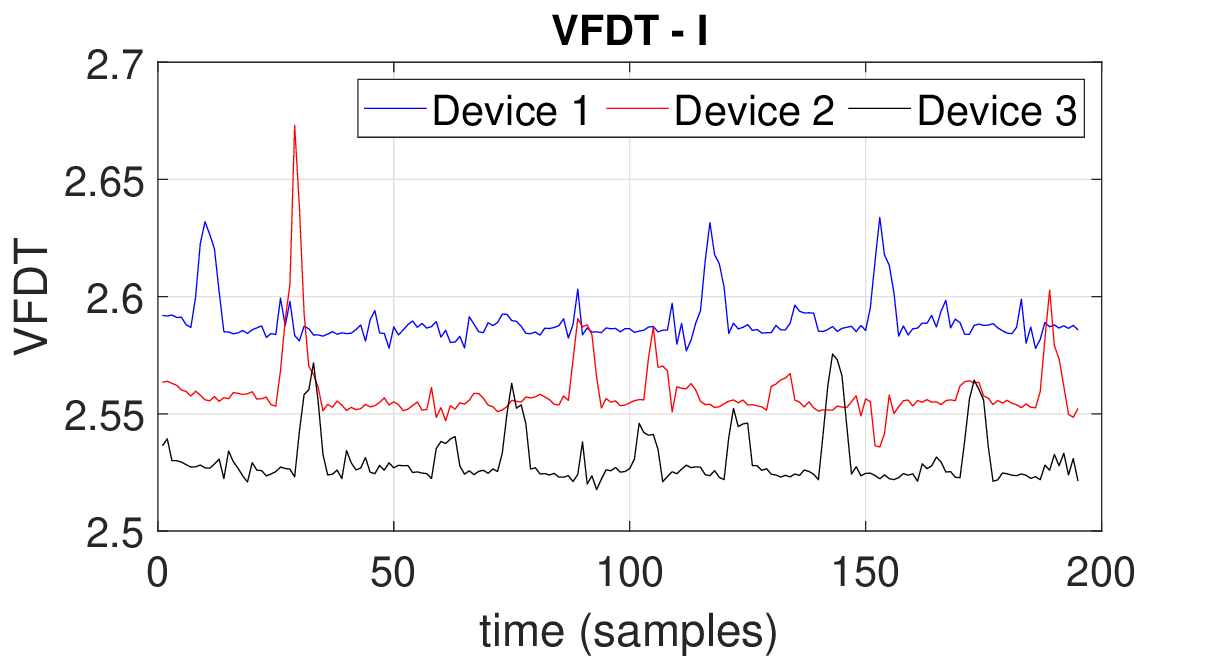}
    \label{fig:lora}}
    \hspace{-0.2in}
\subfloat[WiFi]{
    \includegraphics[width=0.47\columnwidth]{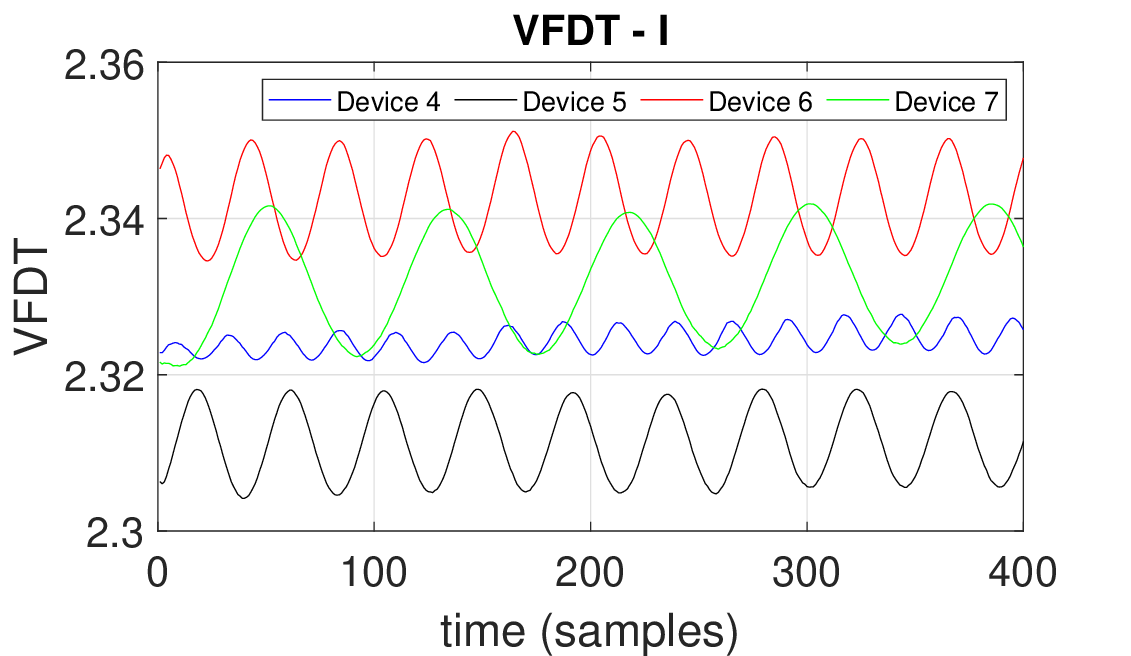}
    \label{fig:wifi}}
\caption{VFDT of the I signal component of Pycom devices.}
\label{fig:vfdt-real}
\end{figure}

Here, we repeat the same experimental setup above except that the WiFi transmission mode is enabled instead of the LoRa mode. For this, we collected IQ data also from 10 different WiFi-enabled LoPy devices, transmitting at the 2.4GHz band. The transmissions were sampled at 45MS/s. Each Pycom device was placed 1m away from the receiver in an indoor environment and transmitted data was collected for 2 minutes. Plots of the calculated VFDT values for the I component of 4 different Pycom devices are shown in Fig. \ref{fig:wifi}. Again, observe that the VFDT values of all 4 devices are consistent across the entire frame and have values that are well distinct across the different devices. 
%
%
One interesting result of note is how each device's VFDT seems to oscillate and at a certain frequency. The oscillating behavior is likely due to the presence of carrier frequency offset between the transmitter and the USRP receiver, which differs from one device to another, resulting in different frequency of the sinusoidal shape~\cite{elmaghbub2023eps}.

\section{Hardware Impairments and Their Impact on the VFDT of the Received IQ Signals}
\label{sec:impairments}
A wireless device's RF transmitter has undesired hardware impairments and variations that result from the inherent non-ideal manufacturing process. 
These impairments vary across devices, resulting in every device having a unique set of impairments, which serve as distinctive signatures that can be extracted and used to uniquely identify the device. While this fingerprint or signature is embedded in the transmitted RF signal, as explained in the introduction section, when the raw RF data is used to train a deep learning model, the embedded fingerprint is typically not captured well enough to correctly classify a device across different domains. Thus, the goal of using multifractal analysis and the proposed VFDT is to allow deep learning models to better extract device-specific fingerprints so as to maintain performance consistency across changing domains. 

\subsection{Device Hardware Impairments}
\begin{figure}[t]
\centering
\includegraphics[width=0.95\columnwidth]{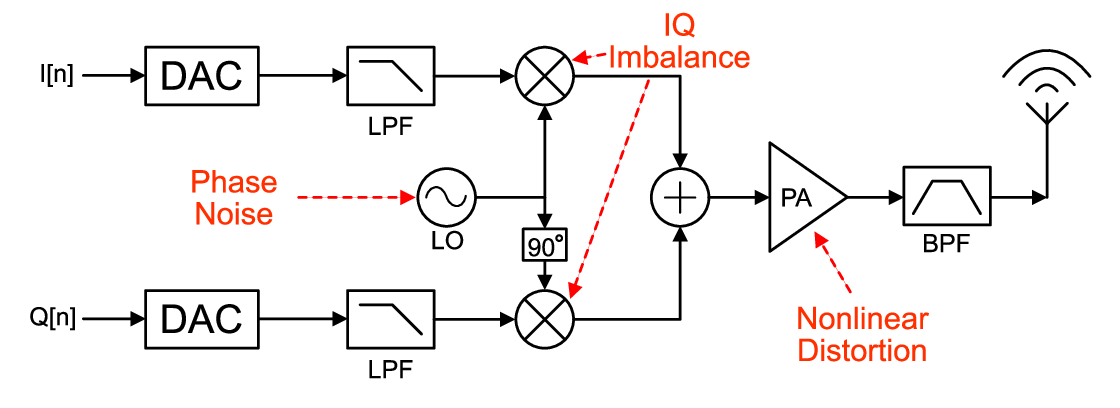}
\vspace{-0.05in}
\caption{Basic RF transmitter front end.}
\label{fig:rf front end}
\vspace{-0.2in}
\end{figure}

As shown in Fig.~\ref{fig:rf front end}, an RF transceiver consists of various hardware components, including digital-to-analog converters (DACs), low-pass filters (LPF), local oscillators (LO), mixers, phase shifters, and power amplifiers (PA). 
Each component comes with undesired hardware impairments that manifest themselves in various forms of distortion, including IQ imbalance, DC offset, phase noise, carrier frequency offset, and power amplifier nonlinearity, and that vary across devices, resulting in each device having a unique set of distinctive fingerprints~\cite{elmaghbub2020leveraging}. While this fingerprint or signature is embedded in the transmitted RF signal, as explained in the introduction section, when the raw RF data is used to train a deep learning model, the embedded fingerprint is typically not captured well enough to correctly classify a device across different domains. Thus, the goal of using multifractal analysis and the proposed VFDT is to allow deep learning models to better extract device-specific fingerprints so as to maintain performance consistency across changing domains.

In Sec.~\ref{subsec:sep}, we experimentally showed the effectiveness and ability of the proposed VFDT representation in capturing device-specific features and in separating among different devices. However, in these real, device-generated signals, it is the aggregation of all the impairments that is captured via VFDT. As such, only the overall effect of the impairments as a whole was analyzed with VFDT. Given that each type of hardware impairments makes its own contribution to the larger overall signal distortion, in this section, we turn our attention to studying the effect of each impairment on the VFDT behavior (and hence on the device fingerprint). Since it is not possible to vary and adjust the value of a hardware impairment on a real device, here we rely on simulation to do so.

\subsection{VFDT Separability Under Individual Impairments}
We studied three key impairments:
PA nonlinearity, IQ imbalance, and phase noise. 
All of the simulations in this section are performed using MATLAB's predefined impairment models.
For each impairment, a random 16,000 bit payload of data is generated. 
Then the 4-QAM modulation is used to digitally modulate the signal, yielding the two in-phase (I) and the quadrature (Q) signal components. The resulting IQ signal is then passed through the specific model for a given impairment. Finally, the received IQ signal output is sampled and analyzed using VFDT. Both the I and Q signals are analyzed individually along with the complex magnitude and phase.

\begin{figure}[t]
\centering
\subfloat[VFDT of I component]{
    \includegraphics[width=0.48\columnwidth]{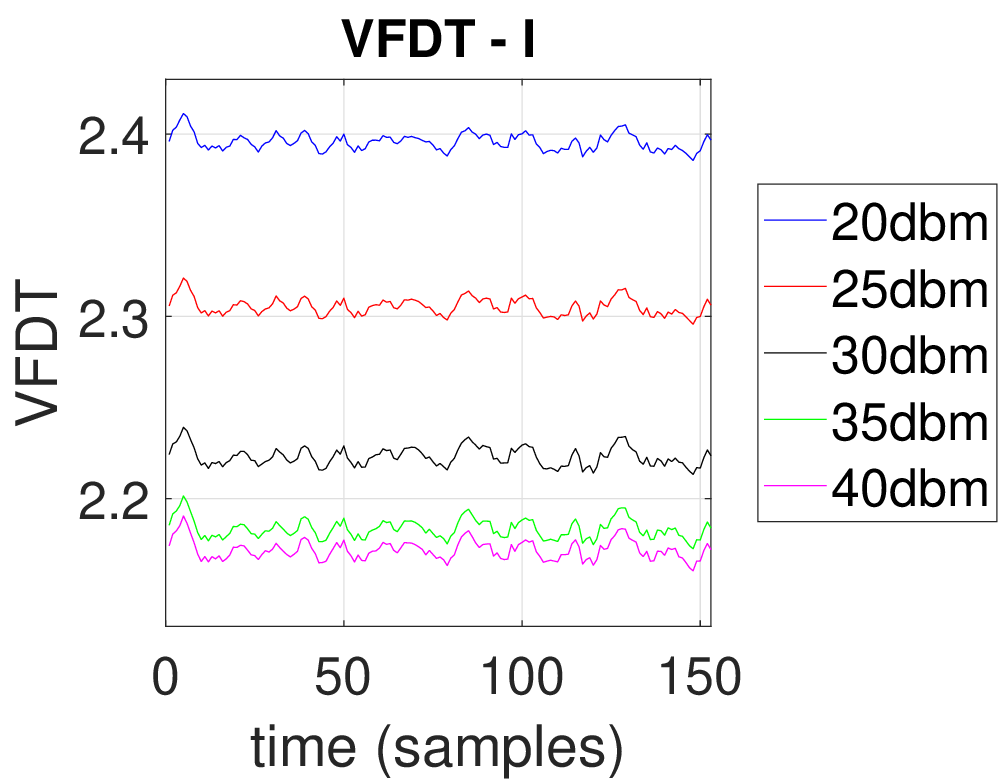}
    \label{fig:pa i}}
\subfloat[VFDT of Q component]{
    \includegraphics[width=0.48\columnwidth]{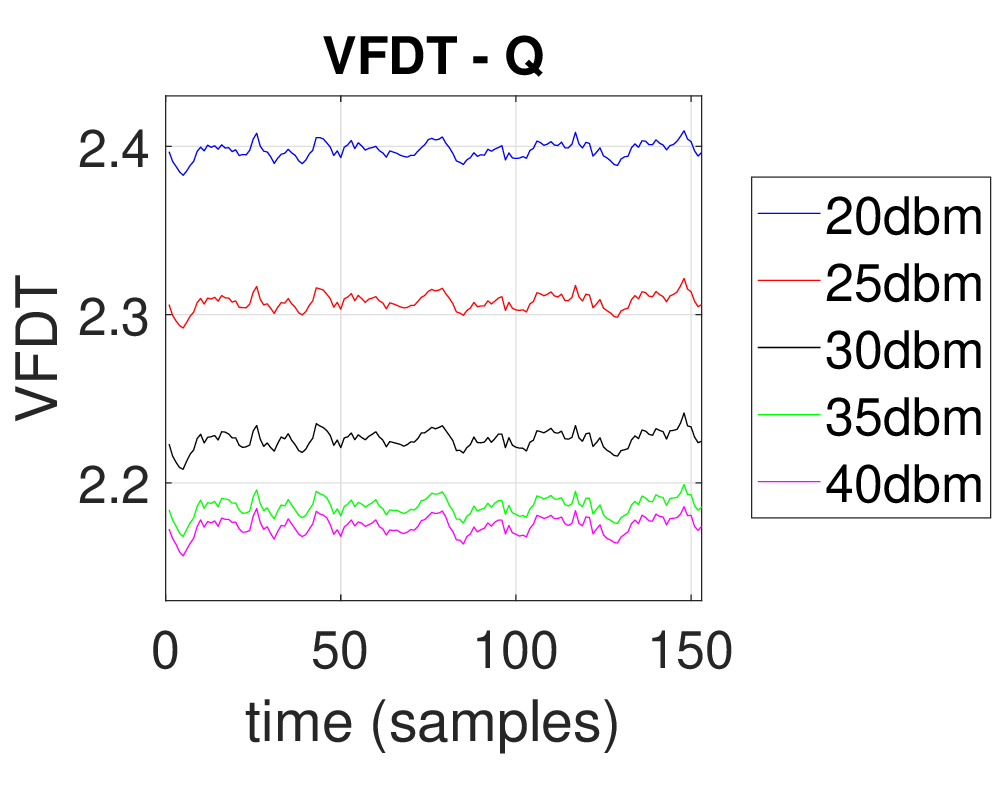}
    \label{fig:pa q}}

\subfloat[Average VFDT across different distortions]{
    \centering
    \includegraphics[width=0.85\columnwidth]{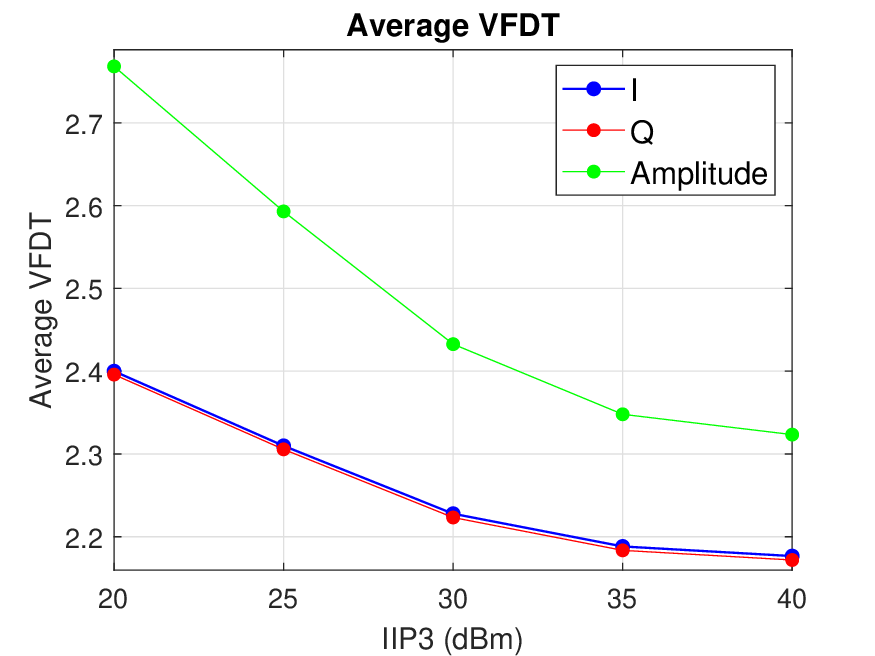}
    \label{fig:pa average}}
\caption{Impact of PA Nonlinearity Distortion on the VFDT of the IQ signals under varying IIP3 values.}
\label{fig:pa sim}
\end{figure}

\subsubsection{Power Amplifier (PA) Nonlinearity Distortion}
PAs amplify the power of the modulated RF signal prior to its transmission on the antenna. While PAs are ideally linear, operating in said region consumes a large amount of power. Thus, in order to run efficiently, they often operate in their nonlinear region, causing a nonlinearity distortion that is mainly a result of the amplitude and phase output responses due to changes in the input signal. These responses, called amplitude-to-amplitude (AM-AM) and amplitude-to-phase (AM-PM) conversions, can be modeled as a complex power series \cite{1231065}. When the PA is linear, only the first order coefficient contributes to the output, but if it is nonlinear, other coefficients contribute to the signal distortion too. Typically, the even order coefficients cancel out leaving the third order coefficient as the main cause of distortion. We use MATLAB's cubic polynomial memoryless nonlinear model to simulate varying degrees of PA nonlinear distortions and analyze their resulting VFDT values.

In order to vary the amount of PA nonlinearity, we change the third order intercept coefficient (IIP3) parameter, which is a measure of the third order distortion described above. After passing the input signal through a given amount of distortion, the VFDT of the output is calculated. We considered 5 different distortion levels, with IIP3 being set to 20dBm, 25dBm, 30dBm, 35dBm, and 40dBm. The VFDT is calculated on the I and Q components of the resulting signal along with the complex amplitude and phase. Finally, the average VFDT of each instance is also plotted across the different levels of distortion. The plots of the resulting I and Q components along with the average VFDT are shown in Fig. \ref{fig:pa sim}. It is clear from Figs. \ref{fig:pa i} and~\ref{fig:pa q} that different IIP3 values yield distinct separation in the VFDT representation, demonstrating that using VFDT to capture PA nonliearity distortions provides very distinct device fingerprints that can be used to distinguish devices from each other.
%
%
From Fig. \ref{fig:pa average}, it can also be seen that as the amount of PA nonlinear distortion increases, the VFDT value decreases in a nonlinear fashion. The exact value of the VFDT and the amount of decrease in it is very similar across the I and Q components of the signal. While the exact value of the resulting VFDT of the complex amplitude of the signal is higher than the I and Q components, the rate of change is similar and seems to simply be scaled up by a small factor. The phase plot of the VFDT is not shown in Fig. \ref{fig:pa sim}, but unlike the other signal components, there is no change in the VFDT as the amount of distortion is increased. Overall, this shows that varying levels of the PA impairment can be well captured through the VFDT representation of the I and Q components, suggesting that VFDT could be used for providing separable IQ-based fingerprints of wireless devices.

\subsubsection{IQ Imbalance}

In typical transmitters like the one shown in Fig. \ref{fig:rf front end}, the I and Q signals are both upconverted to the carrier frequency at the same time with two mixers and a 90\textdegree{}  phase shifter for the quadrature signal. If the mixers are ideal and matched, then there is no imbalance between the I and Q components, resulting in a clean complex output signal. However, for real mixers with some mismatch, there will be an imbalance between the I and Q components, resulting in an amplitude and a phase deviation between the I and Q signal components \cite{elmaghbub2020leveraging}, often referred to as IQ imbalance or mismatch. Both the phase and amplitude mismatch result in distortions of the output signal. 

\begin{figure}[t]
\centering
\subfloat[VFDT of I component]{
    \includegraphics[width=0.48\columnwidth]{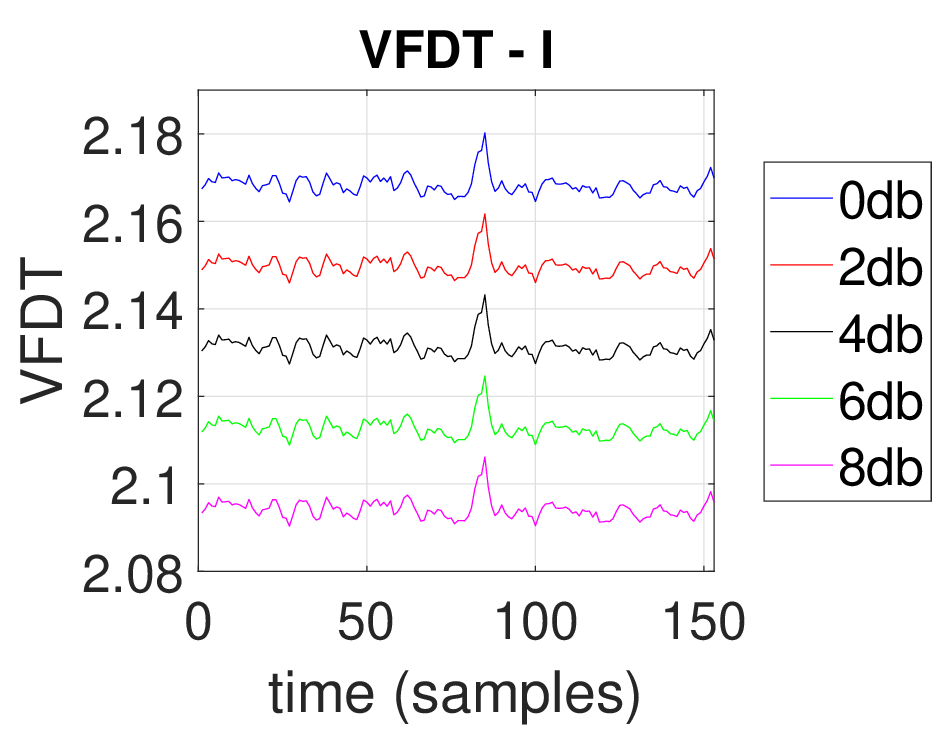}
    \label{fig:iq i}}
\subfloat[VFDT of Q component]{
    \includegraphics[width=0.48\columnwidth]{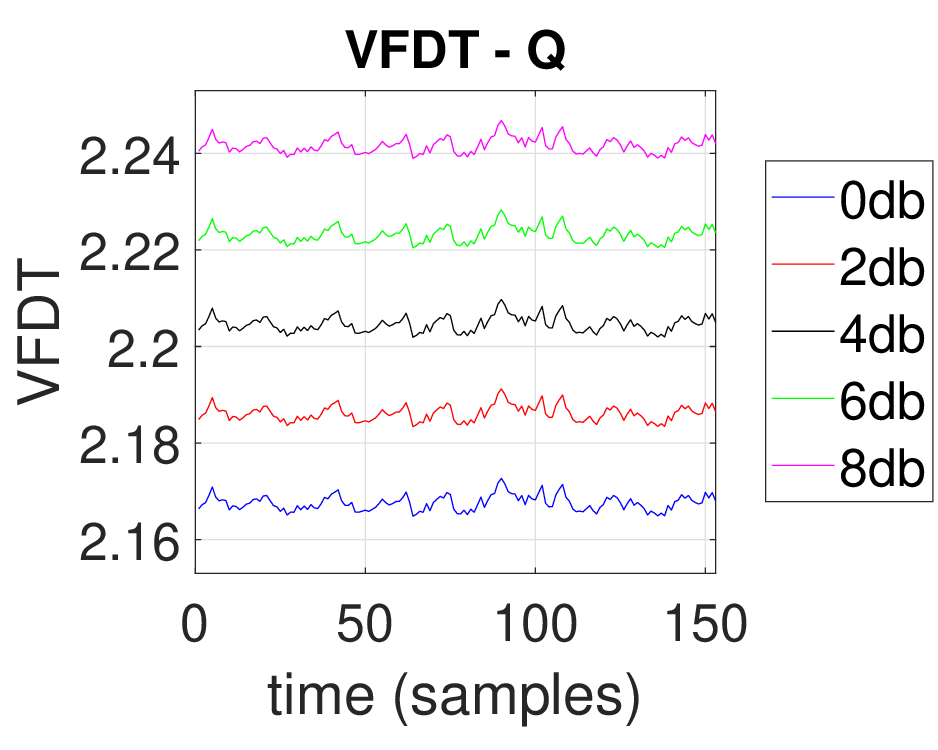}
    \label{fig:iq q}}

\subfloat[Average VFDT across different distortions]{
    \centering
    \includegraphics[width=0.85\columnwidth]{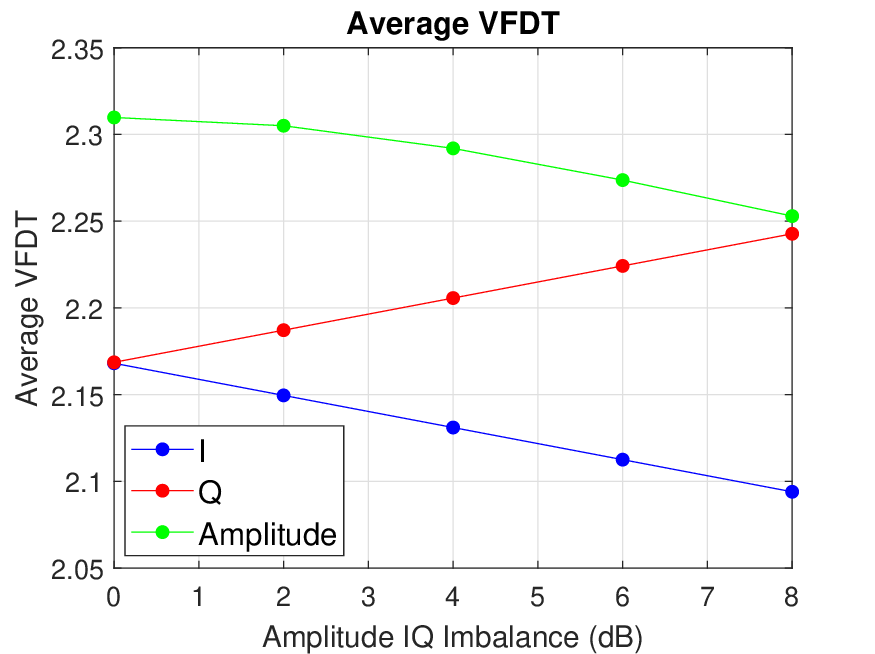}
    \label{fig:iq average}}
\caption{Impact of IQ Imbalance on the VFDT of the IQ signals under varying IQ amplitude imbalance values.}
\label{fig:iq sim}
\end{figure}

For simplicity, only the amplitude imbalance parameter in the model is changed in order to vary the amount of distortion resulting from the IQ imbalance. Again, 5 different amounts of IQ imbalance are used where the amplitude imbalance is set to 0dB, 2dB, 4dB, 6dB, and 8dB. Similar to the previous simulation, the VFDT is taken on the I, Q, complex amplitude and complex phase signals for each scenario. The average VFDT of each instance is plotted across the different levels of impairment. The plots of the I, Q and average VFDT are shown in Fig. \ref{fig:iq sim}. Figs. \ref{fig:iq i} and~\ref{fig:iq q} clearly show direct separation of the VFDT of the I and Q components for differing IQ imbalances. This shows that VFDT is distinct across devices when there is sufficient variation among their IQ imbalances, indicating that VFDT can serve as a good, distinctive device fingerprint. Also, these figures (Figs. \ref{fig:iq i} and~\ref{fig:iq q}) show that the VFDT of the Q signal increases linearly as the amplitude imbalance increases while the VFDT of the I signal decreases linearly as the amplitude imbalance increases. This is interesting as the I and Q components behave differently whereas with the PA distortion both components were affected in the same way. This result makes sense as the model for IQ imbalance assumes an equal and opposite mismatch in amplitude between I and Q signals. The other result of note is that the VFDT of the complex amplitude does not follow either I or Q component, but instead it decreases in a nonlinear fashion as the distortion increases. This could be due to the combined amplitudes of the I and Q components being smaller when the IQ imbalance is large as one of the components' amplitude will be lower as a result. Finally, similar to PA distortion, the VFDT of the complex phase showed no difference as the IQ imbalance varied. Overall, these results also show that hardware impairments in the mixers can be captured well through the VFDT representations of the I and Q signals, again suggesting that VFDT can be used as a good representation for effectively extracting device-specific fingerprints from the received RF transmissions.

\subsubsection{Phase Noise}

The final hardware impairment that we analyze here is phase noise. Recall that the transmitter's local oscillator (LO) produces the carrier frequency that is used to upconvert the I and Q signals. Ideally, an LO produces a pure sinusoidal wave of a specific frequency with a single impulse at the given frequency representing its power spectrum. However, in real LOs there are random phase fluctuations that cause the actual frequency generated to be drifted from the desired frequency. This results in the power spectrum output of the real LO to spread out beyond either side of the desired impulse \cite{847872}.

\begin{figure}[t]
\centering
\subfloat[VFDT of I component]{
    \includegraphics[width=0.48\columnwidth]{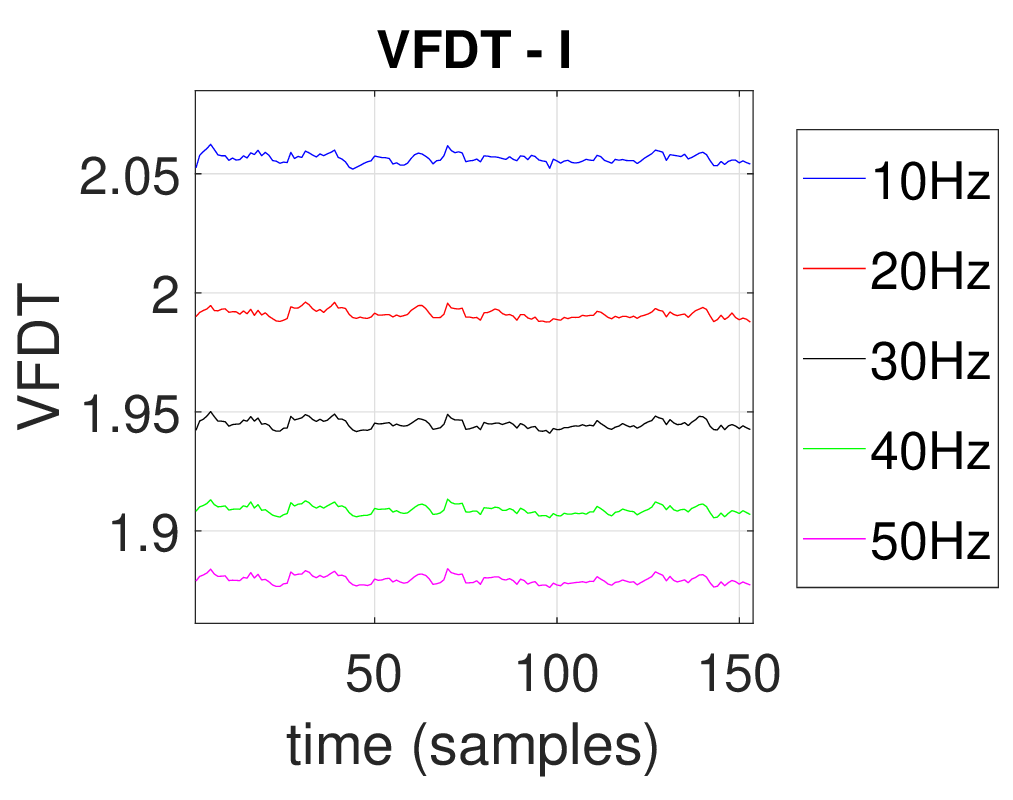}
    \label{fig:phase i}}
\subfloat[VFDT of Q component]{
    \includegraphics[width=0.48\columnwidth]{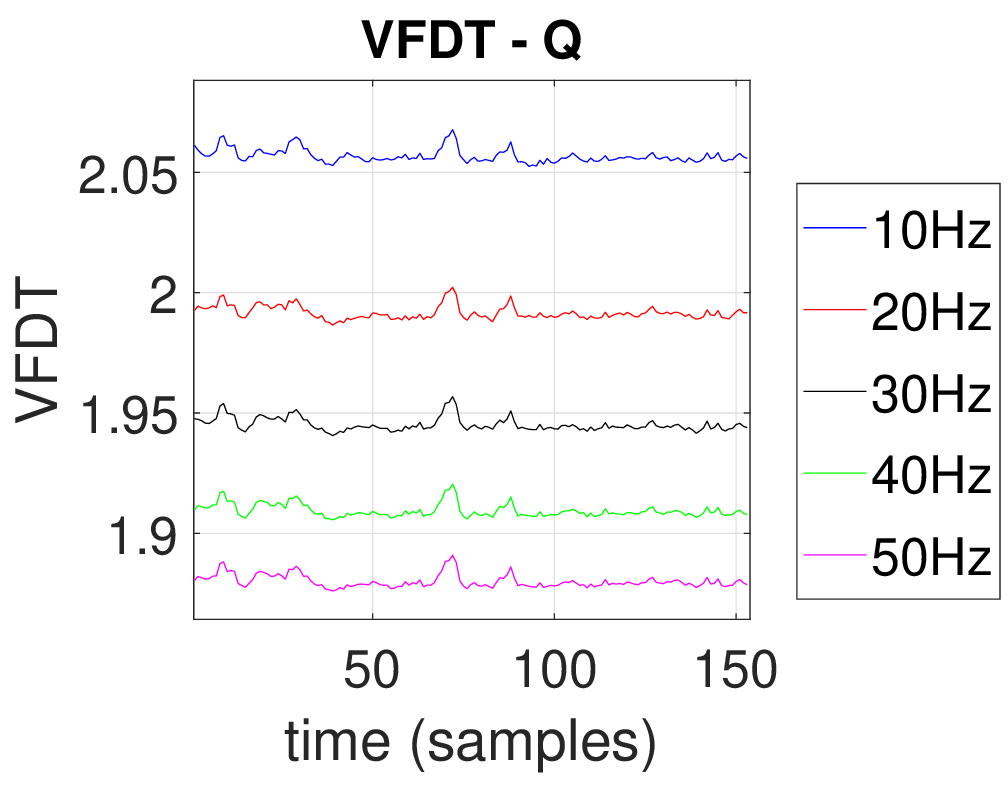}
    \label{fig:phase q}}

\subfloat[Average VFDT across different distortions]{
    \centering
    \includegraphics[width=0.85\columnwidth]{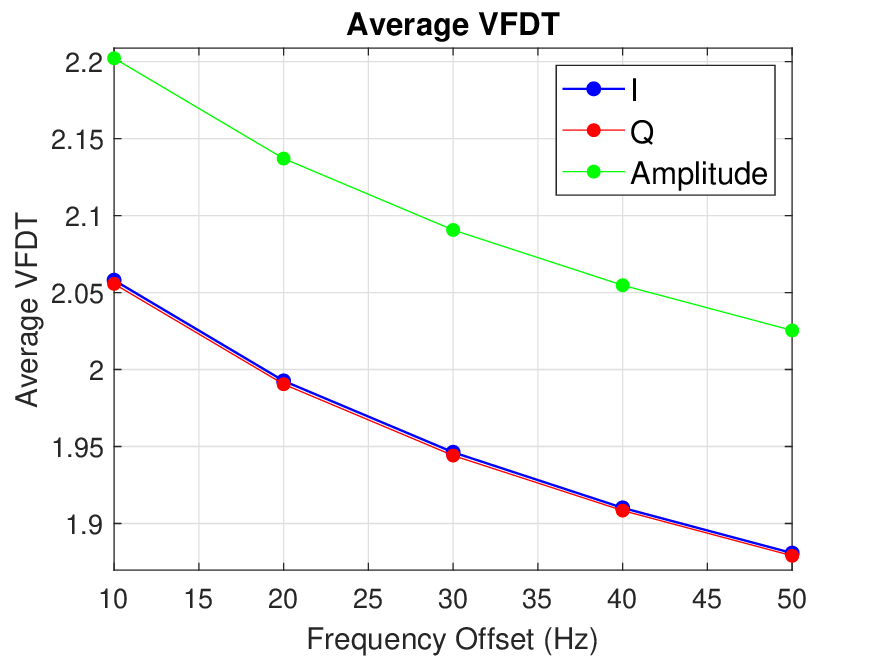}
    \label{fig:phase average}}
\caption{Impact of Phase Noise on the VFDT of the IQ signals under varying phase noise levels.}
\label{fig:phase sim}
\end{figure}

We used the built-in MATLAB model for phase noise, which implements filtered Gaussian noise to model frequency variations. The frequency offset parameter which determines the maximum frequency offset is changed in order to vary the amount of phase noise distortion. We considered 5 different phase noise levels, with the maximum frequency offset being set to 10Hz, 20Hz, 30Hz, 40Hz, and 50Hz. VFDT is then taken on the resulting I, Q, complex amplitude, and complex phase signals. The average VFDT is also plotted across different levels of distortion. The plots of the I, Q and average VFDT are shown in Fig. \ref{fig:phase sim}. Once again, Figs. \ref{fig:phase i} and~\ref{fig:phase q} show that as the amount of phase noise varies, there is clear separation between the VFDT. This implies that the VFDT of different devices would be distinct with enough variation in the impairment, leading to the VFDT being a good device fingerprint. Also, from Figs. \ref{fig:phase i} and~\ref{fig:phase q}, it can be seen that the VFDT of both the I and Q signals decreases as the maximum frequency offset increases. The rate of decrease is not quite linear but consistent across both components. It can also be seen in Fig. \ref{fig:phase average} that the VFDT of the complex amplitude seems to decrease at the same rate as the I and Q signals but is shifted up slightly. While not plotted, the VFDT of the complex phase does not vary at all when the amount of phase noise is changed. Overall, we again show that the LO impairments can be well detected through VFDT representations, and thus can serve well the purpose of extracting device fingerprints from RF transmissions.

\section{The Proposed Classification Method}
\label{sec:proposed}
\begin{figure}[t]
\centering
\includegraphics[width=0.9\columnwidth]{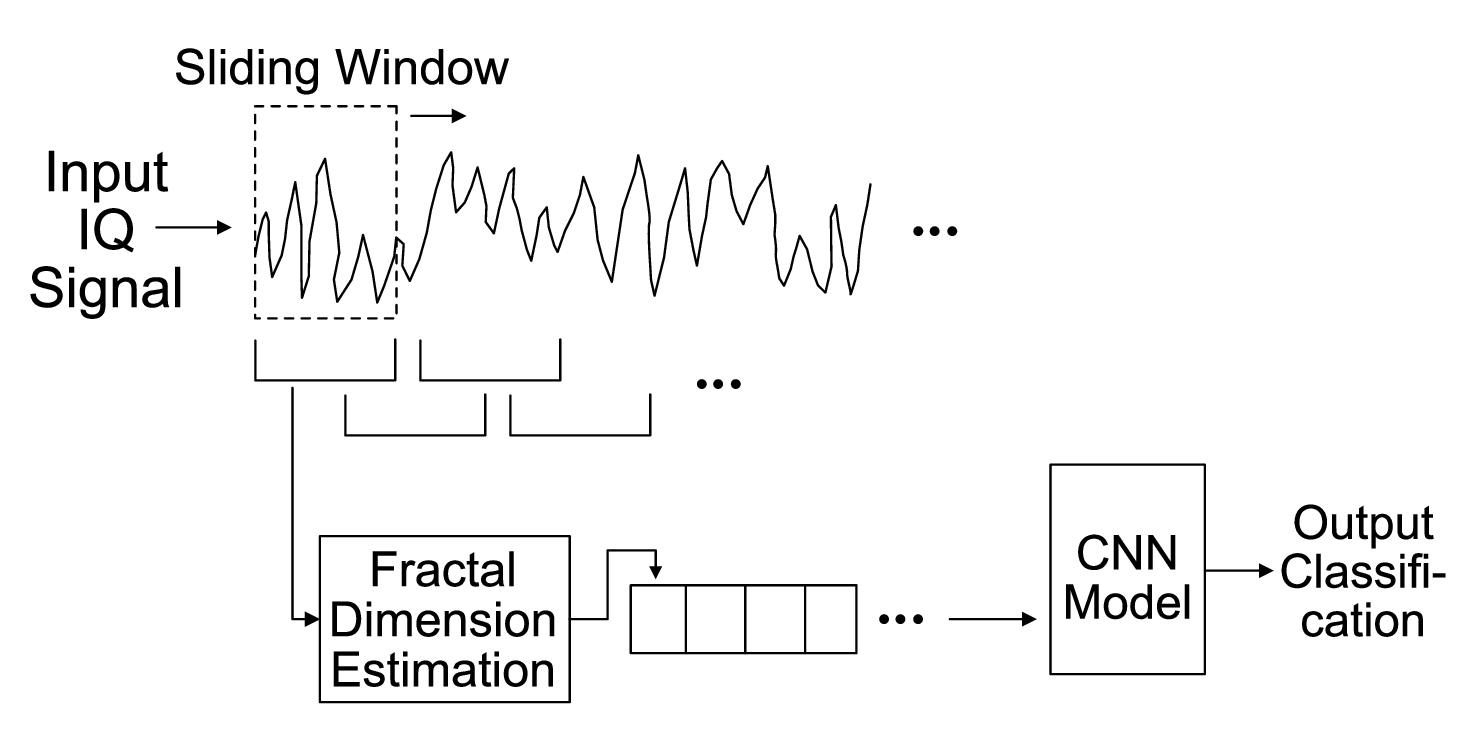}
\caption{Overview of the proposed method.}
\label{fig:proposed method}
\end{figure}

Fig. \ref{fig:proposed method} gives a top-level overview of the proposed method. The goal of the method is to process and extract the fundamental hardware impairments using the VFDT as the input representation before being fed into the deep learning model. A device's RF transmissions are captured by the receiver and split into its I (in-phase) and Q (quadrature) components. The VFDT of both the I and Q component signals are then computed by calculating the variance fractal dimension for every window or segment of the signal using Eq.~\eqref{eg:vfdt}. For each window, the fractal dimension is estimated by first finding the statistical variance of the signal amplitude across all samples in the given window. Next, the log of the variance is taken and divided by the log of the total number of samples in the given window. Finally, the resulting value is scaled and offset to match the definition of the fractal dimension. The computation starts at the beginning of the signal and the window is then shifted by a set predefined amount until the end of the signal is reached. The computed VFDT is a sequence of values representing the variance fractal dimension along different points of the input signal. These two VFDT sequences for the I and Q components are then fed to a convolution neural network (CNN)-based deep learning model, which outputs the final device classification. 

\begin{figure}[t]
\centering
\includegraphics[width=0.65\columnwidth]{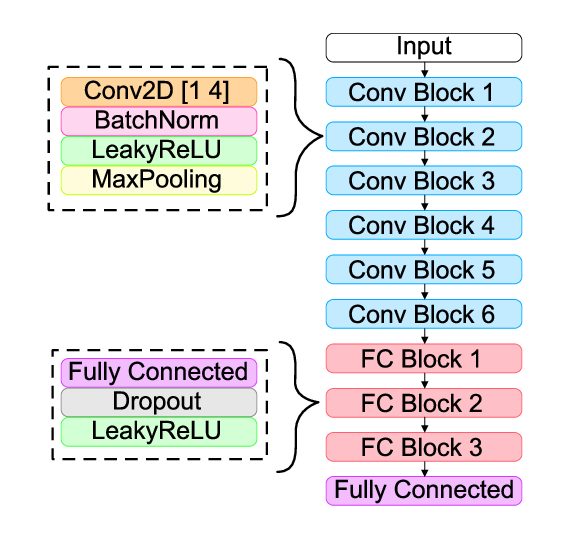}
\caption{CNN architecture for the used deep learning model.}
\label{fig:CNN}
\end{figure}

The CNN architecture used for the deep learning model is based off of the CNN model described and used in \cite{elmaghbub2021Lora}. Its top level architecture is shown in Fig. \ref{fig:CNN}. The CNN is implemented using the "PyTorch" library which is based on the Python programming language. The base architecture consists first of 6 convolutional blocks which are all made up of 4 different layers. The first layer is a 2D convolutional layer, followed by a batch normalization, leaky ReLU, and max-pooling layer, in that order. After the convolutional blocks, there is a sequence of 3 fully connected blocks used to prevent overfitting and help format the network to have the appropriate final outputs. Each fully connected block contains 3 actual layers. The first is a true fully connected linear layer, followed by a dropout and leaky ReLU layers. Finally, there is a single fully connected linear layer to obtain the proper number of output nodes to classify the corresponding number of devices. The CNN takes in a pair of vectors as input. The input vectors are the resulting VFDTs of the I and Q samples, respectively. Separate VFDTs are taken for the I samples and the Q samples and the resulting vectors are fed as inputs to the CNN in the form of 2x1024.

\section{Testbed and Dataset Collection}
\label{sec:testbed}
\begin{figure}[t]
\centering
\subfloat[B210 receiver]{
    \includegraphics[width=.25\columnwidth]{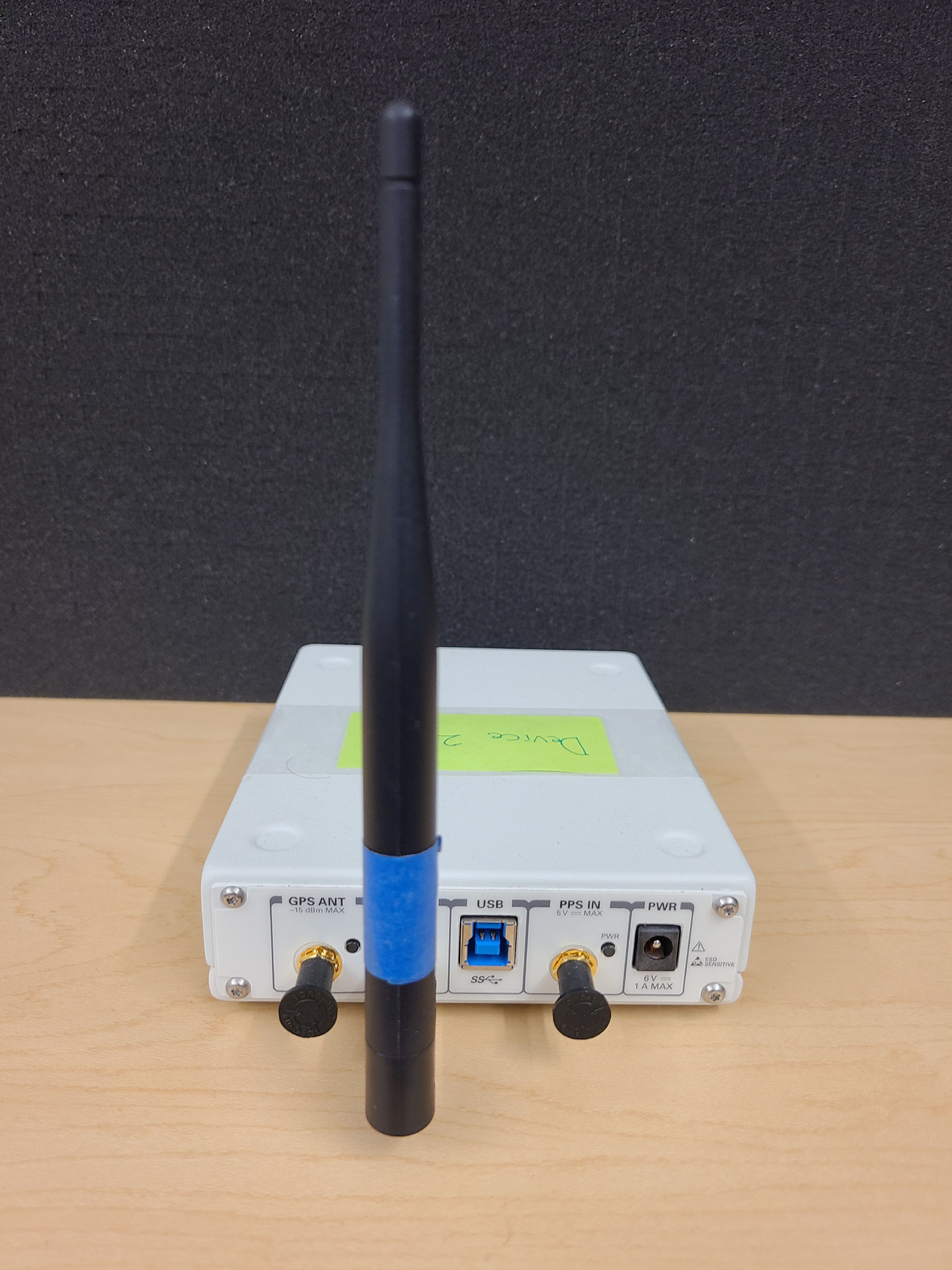}
    \label{fig:ursp}}
\subfloat[1 Lopy and 1 Fipy transmitters]{
    \includegraphics[width=.56\columnwidth]{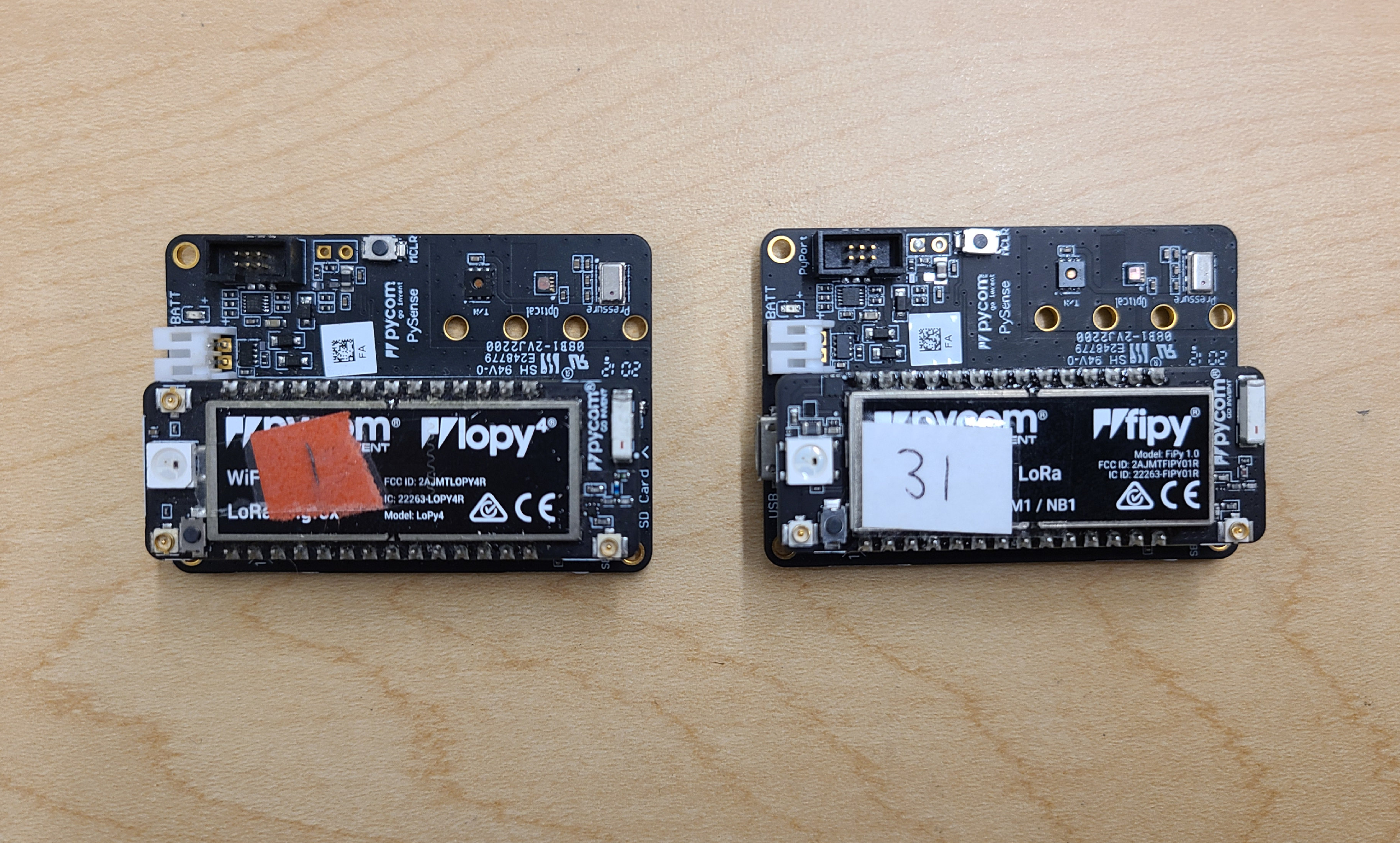}
    \label{fig:pycom}}
\caption{Testbed hardware}
\label{fig:testbed_devices}
\end{figure}

To evaluate VFDT, we collected multiple datasets, each collected from 30 different Pycom devices tested across 5 different locations. These 30 Pycom devices (see Fig.~\ref{fig:pycom}) are made up of 13 Lopy4 boards and 17 Fipy boards, which are IoT boards equipped with a programmable ESP32 that are able to transmit with a number of different protocols, including WiFi and LoRa. An Ettus USRP (Universal Software Radio Peripheral) B210 receiver is used to sample and collect the RF data in the form of raw IQ values through GNURadio (see Fig.~\ref{fig:ursp}).

The first 3 locations tested are at set distances away from the receiver, all within line of sight of the antenna: Location 1 is 1 meter away, Location 2 is 2 meters away, and Location 3 is 3 meters away. Two other considered locations are selected at random from a set of 10 predetermined locations across the testbed area. These locations vary in distance and angle away from the receiver to vary the channel conditions as much as possible. Later in Sec.~\ref{sec:results}, these 2 random locations are referred to as "Rand 1" and "Rand 2". 

Each device is powered on given an initial 20 minute warm up period before data collection begins to ensure hardware stabilization~\cite{elmaghbub2023eps}. Each device is then recorded for 2 minutes continuously beginning with Location 1, followed by Locations 2 and 3, then by Rand1 and Rand2. Each device's ESP32 microcontroller is programmed with the same code to transmit the same message repeatedly. Devices are set to use the WiFi protocol at 2.412GHz with a bandwidth of 20MHz. The USRP receiver is set to sample at 45MSps with a gain of 20dBm.


\section{Device Classification Results}
\label{sec:results}
The testing classification accuracy is used as the metric for evaluating the robustness of the VFDT-based learning models, which is the percentage of the correctly classified tested inputs out of the total number of tested inputs. All models were trained for 30 epochs with 90\% of the collected data used for training and the remaining 10\% used for testing. We evaluate both domain/location adaptation and scalability to measure the robustness of the proposed VFDT representation.

\subsection{Fingerprinting Adaptation to Location Changes}

\begin{figure}[t]
\centering
\includegraphics[width=\columnwidth]{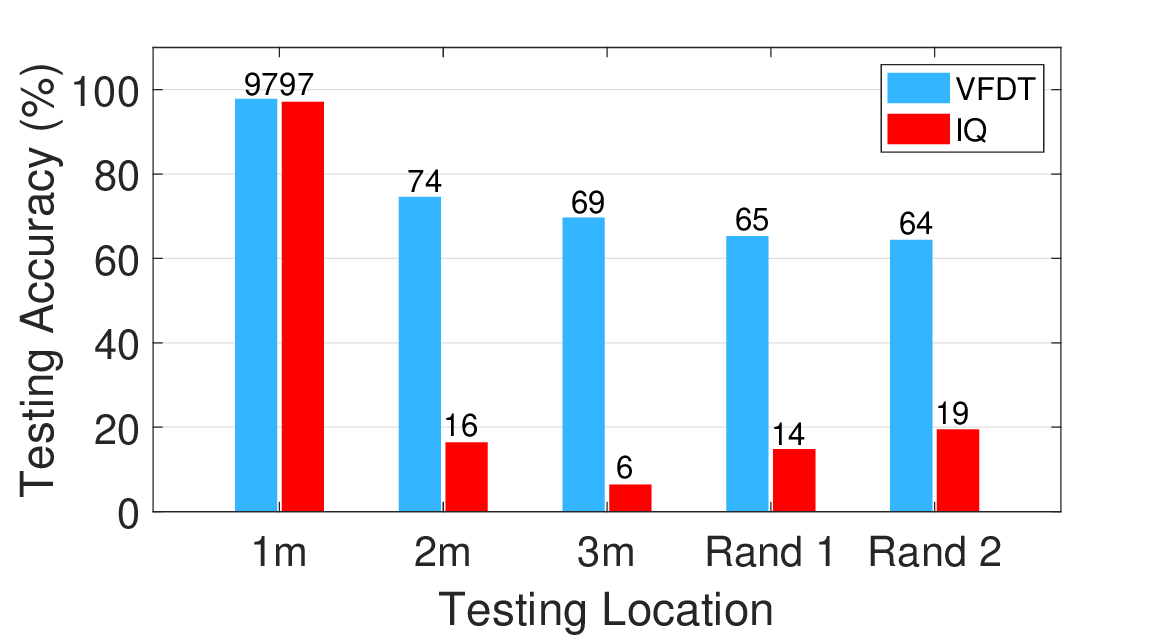}
\caption{Testing accuracy with models trained on Location 1 (1m away) data and tested on data from: Location 1 (1m away), Location 2 (2m away), Location 3 (3m away) and Random Location 1 (Rand 1) and Random Location 2 (Rand 2).}
\label{fig:vfdt vs iq}
\end{figure}

\begin{figure*}[t!]
\centering
\subfloat[IQ: Location 1]{
   \includegraphics[width=0.18\textwidth]{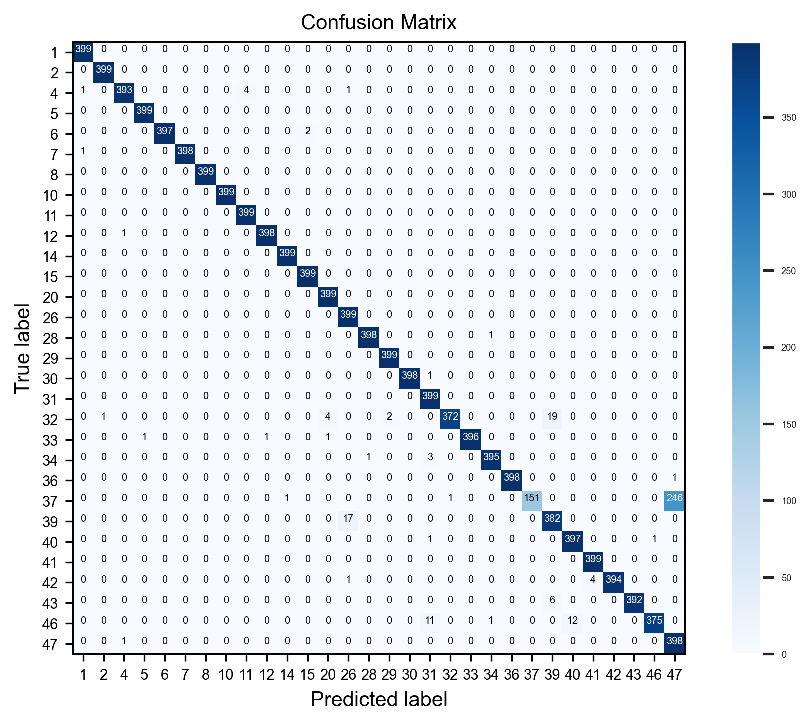}
   \label{subfig:Location 1}}
\subfloat[IQ: Location 2]{
   \includegraphics[width=0.18\textwidth]{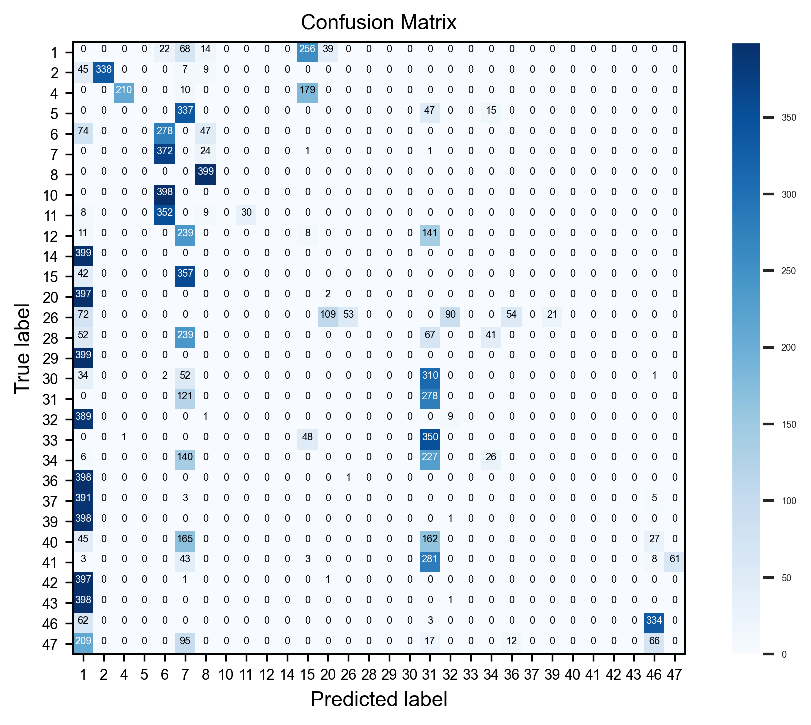}
   \label{subfig:Location 2}}
\subfloat[IQ: Location 3]{
   \includegraphics[width=0.18\textwidth]{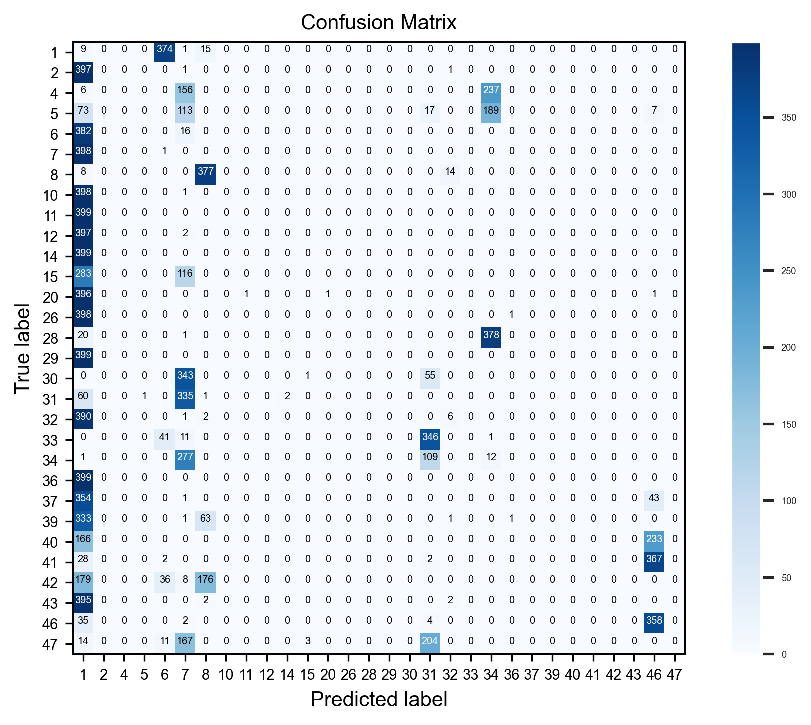}
   \label{subfig:Location 3}}
\subfloat[IQ: Rand1]{
   \includegraphics[width=0.18\textwidth]{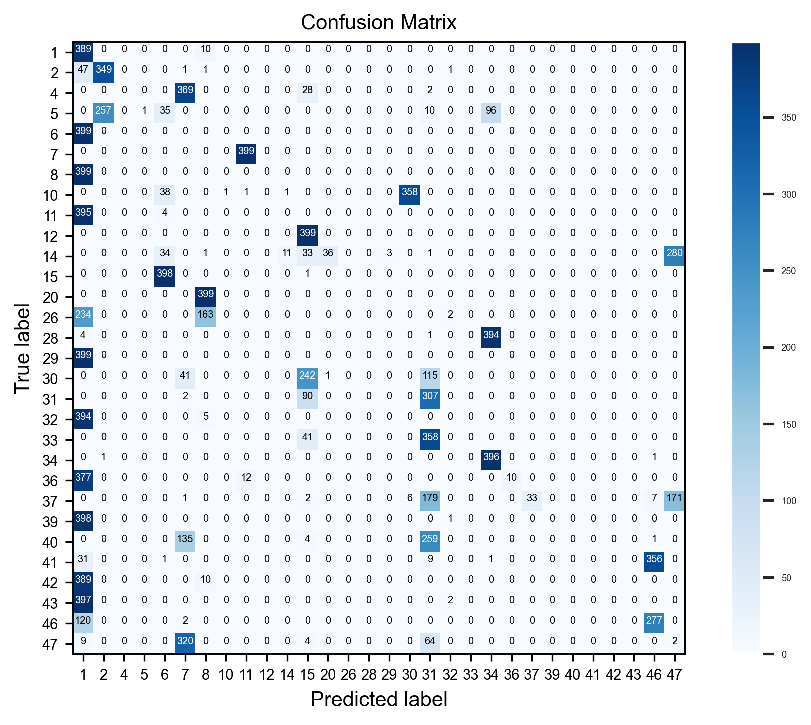}
   \label{subfig:Location 4}}
\subfloat[IQ: Rand2]{
   \includegraphics[width=0.18\textwidth]{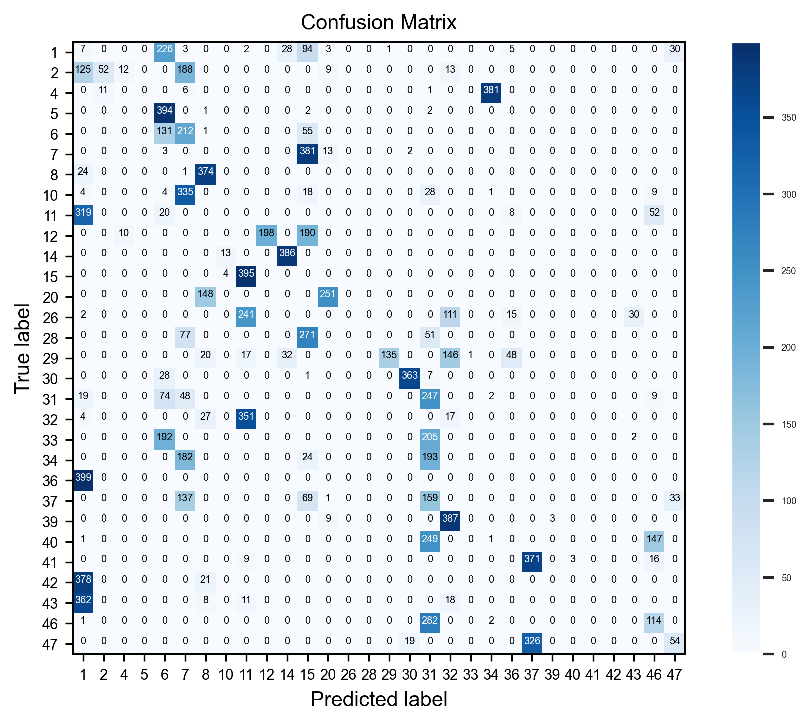}
   \label{subfig:Location 5}}

\subfloat[VFDT: Location 1]{
   \includegraphics[width=0.18\textwidth]{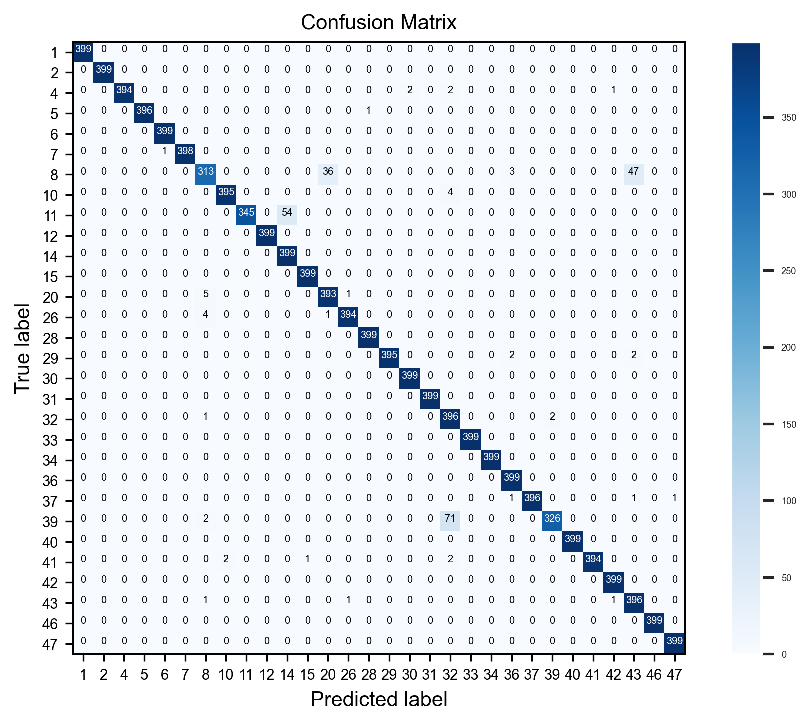}
   \label{subfig:v Location 1}}
\subfloat[VFDT: Location 2]{
   \includegraphics[width=0.18\textwidth]{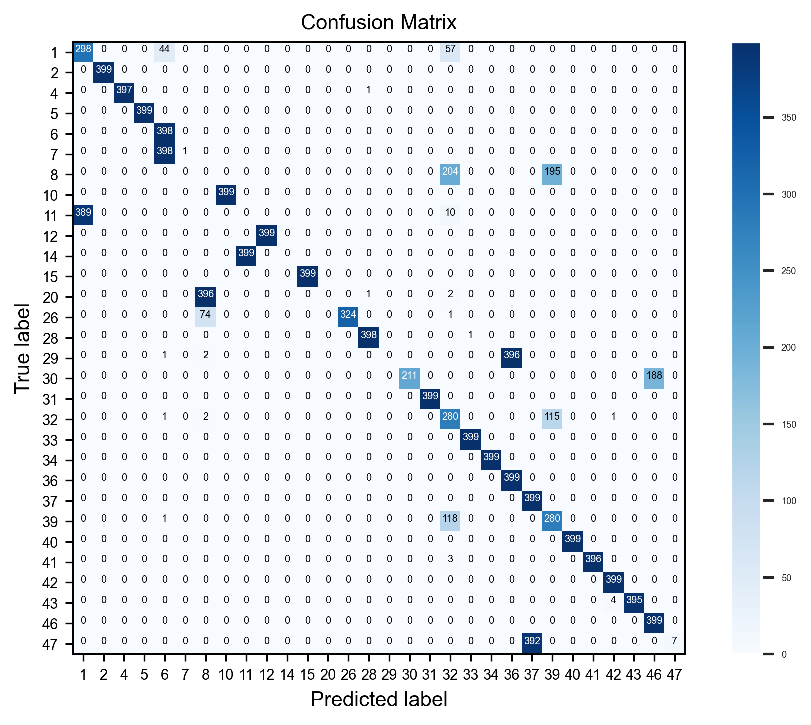}
   \label{subfig:v Location 2}}
\subfloat[VFDT: Location 3]{
   \includegraphics[width=0.18\textwidth]{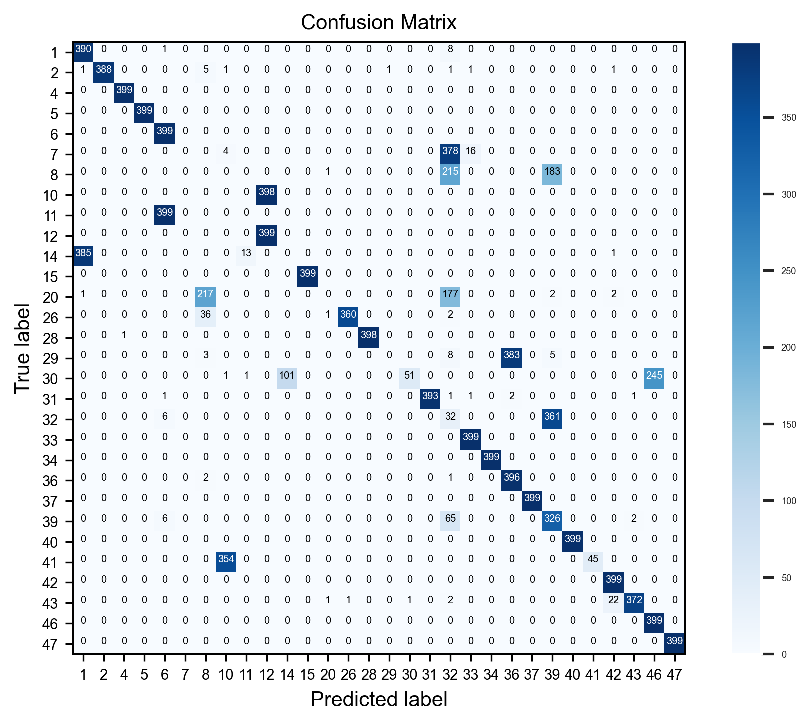}
   \label{subfig:v Location 3}}
\subfloat[VFDT: Rand1]{
   \includegraphics[width=0.18\textwidth]{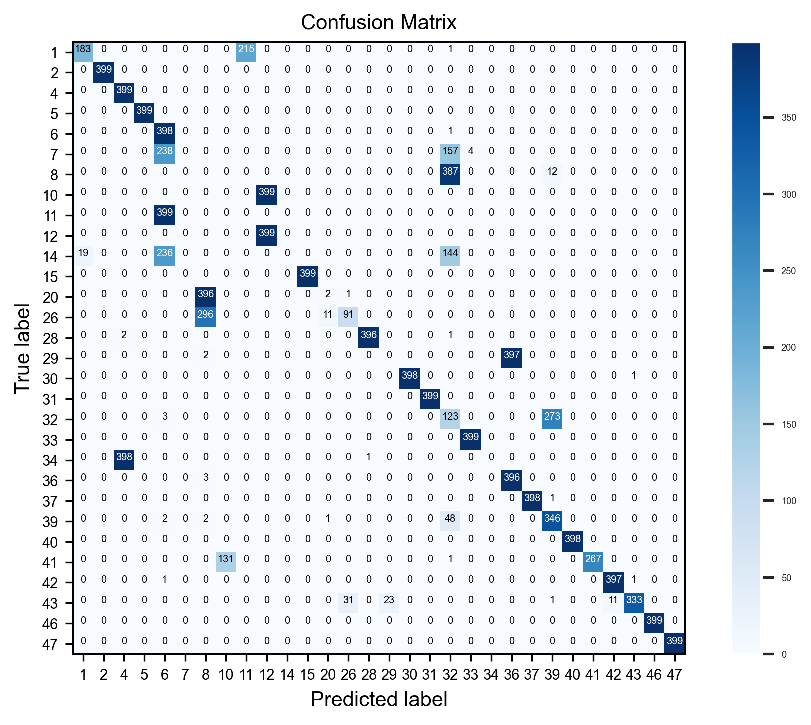}
   \label{subfig:v Location 4}}
\subfloat[VFDT: Rand2]{
   \includegraphics[width=0.18\textwidth]{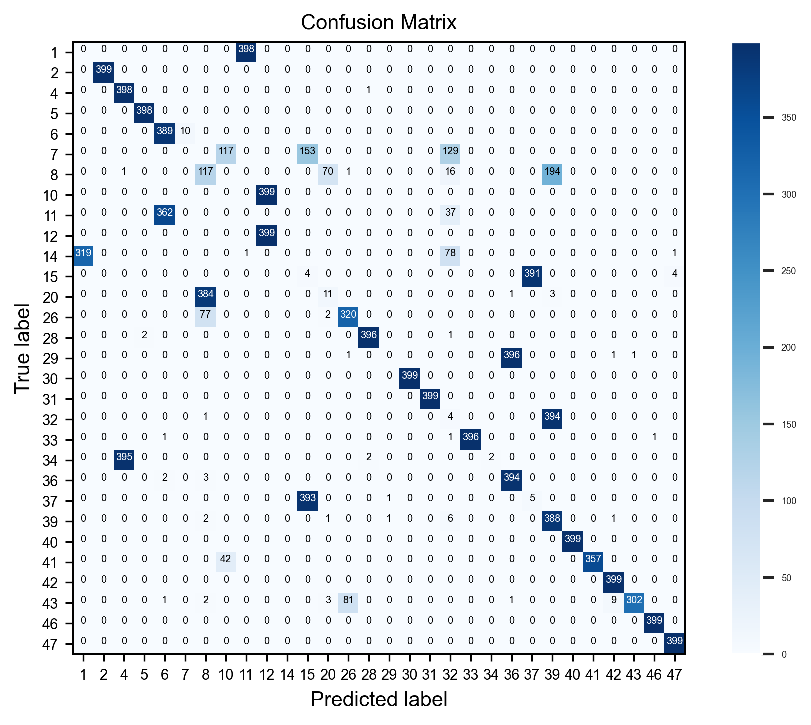}
   \label{subfig:v Location 5}}
\caption{Confusion matrices for 30 tested devices at different locations. Model is trained on Location 1 data.}
\label{fig:matrix}
\end{figure*}

For this evaluation, the proposed VFDT-based deep learning model is trained on VFDT data collected at Location 1, located 1m away from the receiver, and tested on data collected from 5 different locations, including the training location. For a baseline comparison, an identical learning model is trained on raw IQ data from Location 1 and also tested on data from all 5 locations. In all the figures presented in this section, the IQ-based baseline approach is referred to as `IQ', whereas the VFDT-based approach is referred to as `VFDT'.
The testing accuracy for these two approaches is shown in Fig. \ref{fig:vfdt vs iq}, where it can be seen that while both VFDT and IQ achieve similar results when tested on data collected at Location 1 (i.e., when training and testing are done at the same location), VFDT significantly outperforms IQ when tested on data collected from a location different from the training location, Location 1. 
Observe that while the accuracy achieved under IQ drops below 20\% for the other 4 locations, VFDT maintains an accuracy in the high 60\% to low 70\%. Fig. \ref{fig:matrix} sheds more light on the approaches' ability to classify devices across locations. The figure shows that both models classify basically every device correctly when tested at the same location of training. However, the IQ approach's classification becomes very random and seemingly overfit to a single device when tested on data collected at other locations. 
Meanwhile, the VFDT approach is able to classify the majority of devices correctly regardless of the location being tested at. Note that both approaches begin to misidentify devices as the tested location deviates further from the training location, but the drop in accuracy is more significant with the IQ approach than with the VFDT approach. In recap, our results do demonstrate that the proposed VFDT model is generalizable across different locations.

\begin{figure}[t]
\centering
\includegraphics[width=\columnwidth]{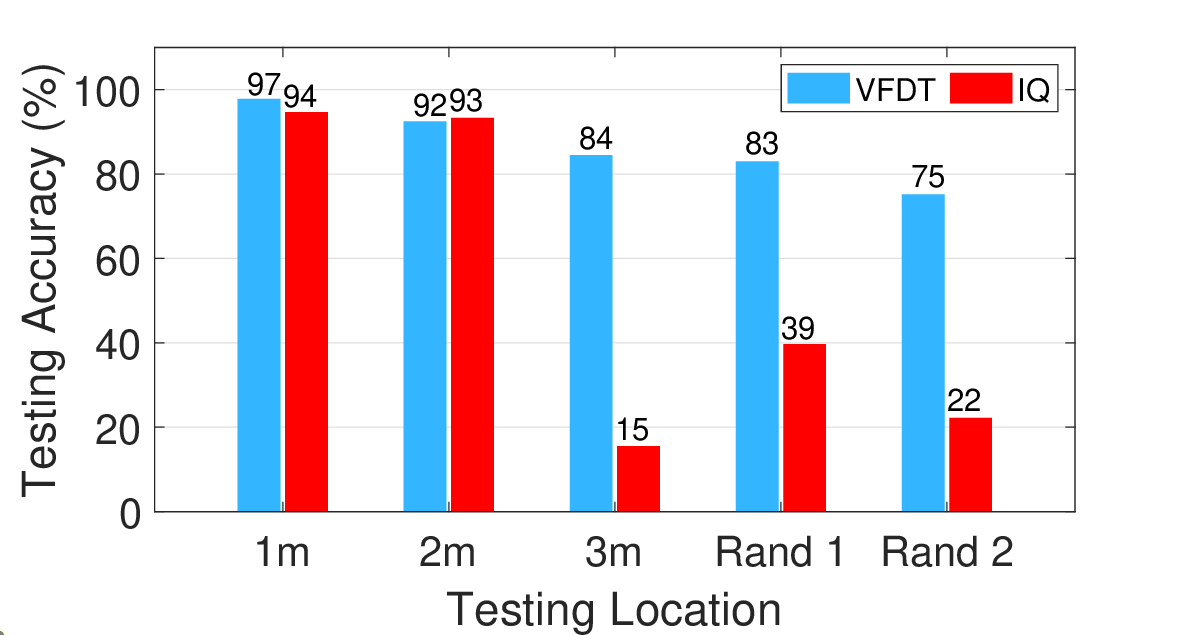}
\caption{Testing accuracy with models trained on mixed data from Location 1 (1m away) and Location 2 (2m away), and tested on data from: Locations 1 (1m away), 2 (2m away), 3 (3m away) and Random Locations 1 (Rand 1) and 2 (Rand 2).}
\label{fig:train12}
\end{figure}

This trend of the VFDT being more consistent and robust across domains than IQ remains present in the second evaluation scenario, where both models are trained on a mixture of data from Location 1 and Location 2 and tested on data collected from one of the 5 considered locations. The overall results are shown in Fig. \ref{fig:train12}, where it can be seen that again VFDT outperforms IQ. The VFDT model achieves greater results when tested on any of the 3 locations (Location 3, Rand 1 or Rand 2) that it was not trained on, once again outperforming the IQ model significantly. 
Comparing the results in Fig.~\ref{fig:train12} with those in Fig.~\ref{fig:vfdt vs iq}, it is worth noting that the IQ model performs slightly better on data taken at Location 3, Rand 1 and Rand 2. For example, when tested on data from Location 3, the accuracy of IQ model jumps from 6\% to 15\%. This is attributed to the benefit of training on mixed data collected from two locations, i.e., Locations 1 and 2.

\begin{figure}
\centering
\includegraphics[width=\columnwidth]{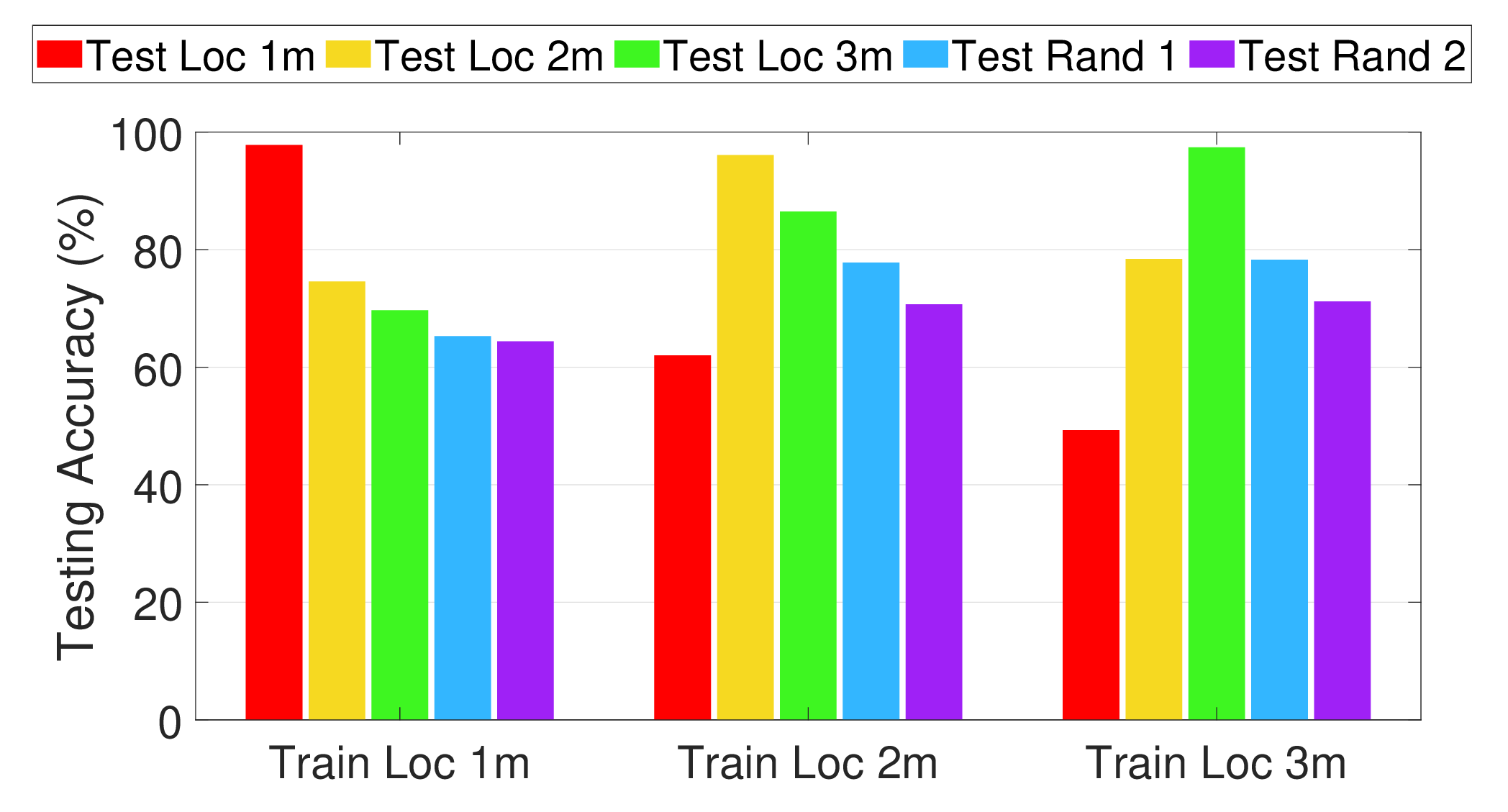}
\caption{Testing accuracy when VFDT model is trained on data collected at different locations.}
\label{fig:train123}
\end{figure}

To ensure performance consistency of the proposed VFDT across other training locations, we also consider a third scenario where the VFDT model is trained on data collected from one of the 2 other two locations, Locations 2 and 3, in addition to the already tested Location 1. 
All models are then also tested on all 5 locations. Fig. \ref{fig:train123} shows that the VFDT model performs relatively well regardless of what location it was trained on. For just about all testing locations that are not a part of training, the testing accuracy is very similar. In all 3 cases, as expected, the model achieves its highest accuracy when tested on the same training location, with a slight drop off for all the other locations. 
Overall, VFDT is able to generalize across different locations by achieving and maintaining decently consistent classification accuracy even under changing locations.

\subsection{Scalability of the VFDT Model with the Number of Devices}

\begin{figure}[t]
\centering
\includegraphics[width=\columnwidth]{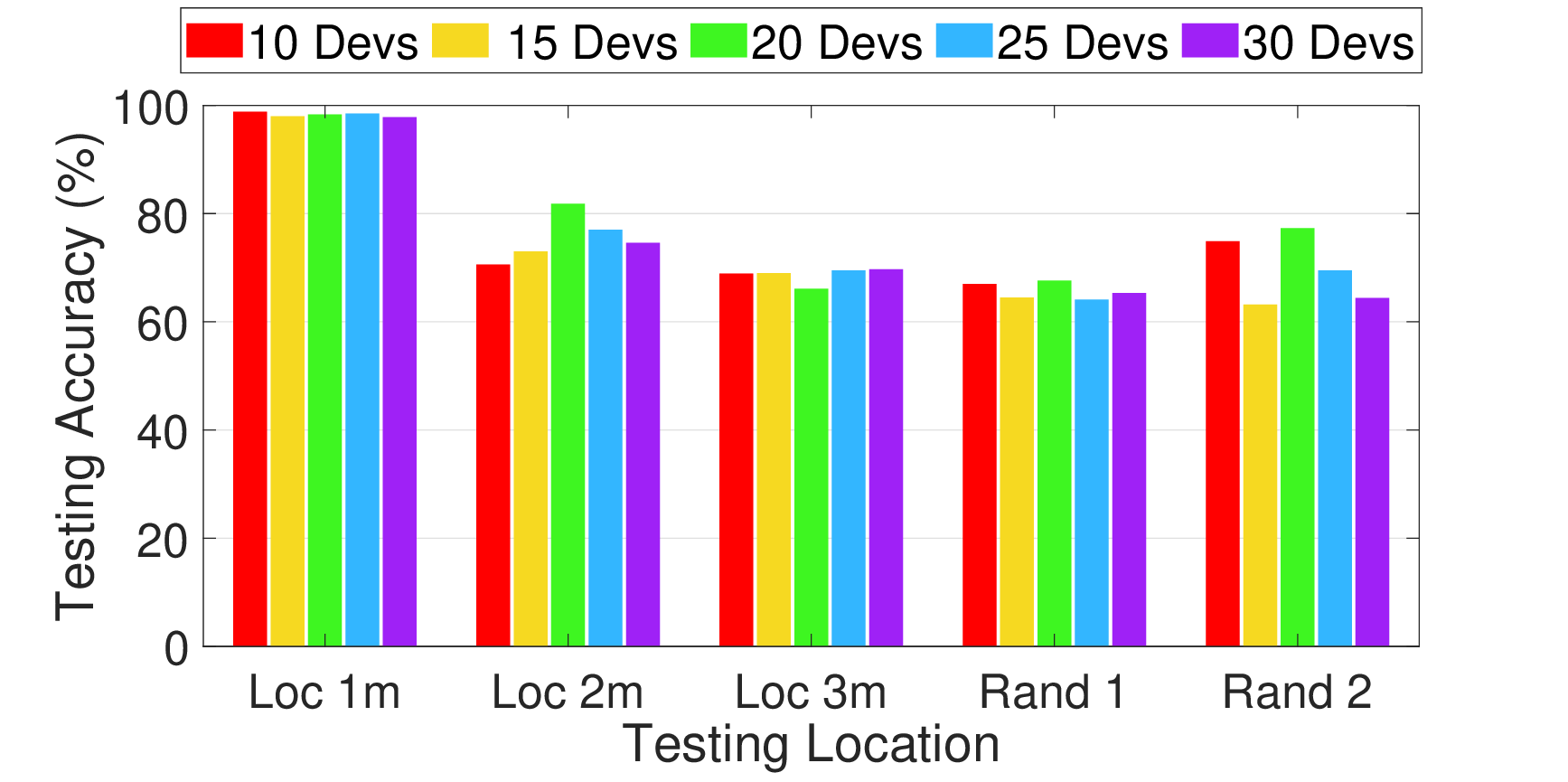}
\caption{Scalability results when models are tested on varying number of devices. All models trained on Location 1 data.}
\label{fig:scale}
\end{figure}

For the final evaluation, we looked at how well VFDT scales with the numbers of devices. 
For this, we trained the models on 5 different subsets of devices, with sizes: 10, 15, 20, 25, and 30 devices. 
For each subset, all models are trained on Location 1 data (again located 1m away from the receiver) and then tested on all other locations. The testing accuracy for all cases is shown in Fig. \ref{fig:scale}. From the figure, it can be seen that there is no significant drop or falloff in accuracy for a given location across all numbers of devices. Whether the VFDT model is classifying 15 or 30 devices, it is able to do so with very similar, consistent performance. Overall, these results indicate that our proposed VFDT representation scales well with the number of devices even under varying locations. This statement is, of course, valid for the sizes that were tested in this work, which we realize that they are still small sizes, and thus further research is needed to look at what happens when considering higher scales.

\section{Conclusion}
\label{sec:conc}
We propose using multifractal analysis through the variance fractal dimension trajectory (VFDT) for extracting separable device signatures from  RF signals using deep learning models. Matlab simulations are performed to analyze how individual hardware impairments affect the VFDT representation vis-a-vis of its ability to separate between devices and fingerprints. It is then demonstrated through the collection of experimental datasets that the proposed VFDT representation of the IQ signals improves the scalability, robustness and generalizability of the deep learning models significantly compared to when using IQ data samples.

\bibliographystyle{IEEEtran}
\bibliography{References}

\end{document}